\documentclass[twocolumn,showpacs,preprintnumbers,amsmath,amssymb,epsfig,floatfix,prx]{revtex4-1}
\usepackage{pdfpages}
\usepackage{pgffor}
\usepackage{etoolbox}
\makeatletter
\patchcmd{\@outputpage@head}{\@ifx{\LS@rot\@undefined}{}{\LS@rot}}{}{}{}
\makeatother

\usepackage{amsmath}
\usepackage{mathtools}
\usepackage{xcolor}
\usepackage{physics}
\DeclareMathAlphabet{\altmathcal}{OMS}{cmsy}{m}{n}
\usepackage{siunitx}
\usepackage{multirow}
\usepackage{amsfonts}
\usepackage{amssymb}
\usepackage{calligra}
\usepackage{calrsfs}
\DeclareMathAlphabet{\mathcalligra}{T1}{calligra}{l}{m}
\usepackage{epsfig}
\usepackage{graphicx}
\usepackage{dcolumn}
\usepackage{color}
\usepackage{natbib}  
\usepackage{hyperref}
\usepackage{breakurl}
\hypersetup{colorlinks=true, citecolor=blue, urlcolor=blue, linkcolor=blue}
\usepackage{bm}
\usepackage{booktabs}
\newcolumntype{C}{>{$}c<{$}}
\AtBeginDocument{
\heavyrulewidth=.08em
\lightrulewidth=.05em
\cmidrulewidth=.03em
\belowrulesep=.65ex
\belowbottomsep=0pt
\aboverulesep=.4ex
\abovetopsep=0pt
\cmidrulesep=\doublerulesep
\cmidrulekern=.5em
\defaultaddspace=.5em
}

\newcolumntype{L}[1]{>{\raggedright\arraybackslash}p{#1}}
\newcolumntype{C}[1]{>{\centering\arraybackslash}p{#1}}
\newcolumntype{R}[1]{>{\raggedleft\arraybackslash}p{#1}}
\newcommand{\C}{{\rm{}{}{}C}}

\newcommand\mybooklet[2][]{
  \pdfximage{#2}
  \pgfmathtruncatemacro\totalpages{\the\pdflastximagepages}
  \pgfmathtruncatemacro\halfpages{ceil(\totalpages/2)}%
  \ifnumodd{\totalpages}{%
    \includepdf[pages={{},1},nup=2x1,#1]{#2}%
    \foreach \pfirst[evaluate=\pfirst as \plast using int(\totalpages+2-\pfirst)]
    in {2,...,\halfpages}{%
      \ifnumodd{\pfirst}%
      {\includepdf[pages={\plast,\pfirst},nup=2x1,#1]{#2}}%
      {\includepdf[pages={\pfirst,\plast},nup=2x1,#1]{#2}}%
    }%
  }{%
    \foreach \pfirst[evaluate=\pfirst as \plast using int(\totalpages+1-\pfirst)]
    in {1,...,\halfpages}{%
      \ifnumodd{\pfirst}%
      {\includepdf[pages={\plast,\pfirst},nup=2x1,#1]{#2}}%
      {\includepdf[pages={\pfirst,\plast},nup=2x1,#1]{#2}}%
    }%
  }%
}

\begin{document}
\title{Long-range anisotropic Heisenberg ferromagnets and electrically tunable ordering}
\author{Chandan K. Singh}
\author{Mukul Kabir}
\email{mukul.kabir@iiserpune.ac.in}
\affiliation{Department of Physics, Indian Institute of Science Education and Research, Pune 411008, India}
\date{\today}

\begin{abstract}
Recent realizations of intrinsic magnetic order in truly two-dimensional materials have opened new avenues in the fundamental knowledge and spintronic applications. Here we develop an anisotropic Heisenberg model with relativistic exchange interactions that are obtained from the first-principles calculations. We demonstrate the crucial importance of magnetic interactions beyond the first-neighbour to qualitatively and quantitatively reproduce the experimental results. Once we ascertain the predictive capacity of the model for chromium trihalides and CrGeTe$_3$, we investigate the feasibility of tuning the magnetic ordering by electrical means in these materials. A remarkable five-fold increase in the ferromagnetic Curie temperature is predicted in monolayer CrI$_3$ within experimentally obtainable hole density. The elusive microscopic mechanism behind the doping-dependent Curie temperature is illustrated. Further, in the present context, the effects of biaxial strain and chemical doping are also addressed. The results should trigger further experimental attention to test the present conclusions.  
\end{abstract}
\maketitle

\section{Introduction}
The magnetic anisotropy arising from the intrinsic spin-orbit coupling breaks the continuous rotational symmetry of spins on a lattice. Consequently, and in apparent contradiction to the Hohenberg-Mermin-Wagner theorem,~\citep{PhysRev.158.383,PhysRevLett.17.1133} the magnetic phase transition in two-dimension (2D) becomes susceptible at a finite temperature.~\citep{nanolett.6b03052,nature22060,nature22391,Gibertini2019} Both single-ion  and exchange anisotropies play decisive roles for  the appearance of long-range magnetic order in 2D, and the long-wavelength excitation becomes gaped.  Ising model description of magnetism is adequate only when the single-ion anisotropy is very strong, and an ordered phase emerges at a finite temperature as predicted by Onsager.~\citep{PhysRev.65.117}  In contrast, the case with weak single-ion anisotropy is somewhat complicated, and an anisotropic Heisenberg model is required, where the spins can freely rotate in the three-dimensional space.~\citep{PhysRevLett.61.2585,Lado_2017,Torelli_2018}  Accordingly, the ordering temperature is considerably suppressed due to spin fluctuations.  
In contrast, for a system with strong in-plane single-ion anisotropy, the magnetism is described by the XY model, and a quasi-long-range magnetic order is established by the Berezinskii-Kosterlitz-Thouless mechanism of topological defect formation.~\citep{Berezinskii1972,Kosterlitz_1973} 

Historically the search for magnetism in reduced dimension remained confined in the ultrathin films~\citep{Poulopoulos_1999,Vaz_2008} and layered magnetic materials with weak interlayer interactions.~\cite{cryst7050121} While the ultrathin films of conventional bulk magnets retain magnetic ordering, their ordering temperature plunges much below the room-temperature in the quasi-2D limit.~\citep{Poulopoulos_1999,Vaz_2008} Further, their application becomes limited owing to the uncontrollable interaction with substrates and difficulties to produce uniform ultrathin layers. To circumvent these drawbacks, the search for van der Waals 2D magnets becomes imperative. However, it was only recently in 2016; the existence of magnetic orderings was discovered in truly 2D materials. Monolayer FePS$_3$ exhibited antiferromagnetic ordering,~\citep{nanolett.6b03052} while ferromagnetism was observed in atomically thin CrGeTe$_3$ and CrI$_3$ in the bilayer and monolayer limits, respectively.~\citep{nature22060,nature22391} These seminal experimental results triggered a new field of research involving 2D magnets and attracted enormous attention due to fundamental physics interest and as well as applications in emergent technologies. Thereafter, a few other insulating and itinerant 2D ferromagnets have been discovered such as CrCl$_3$, CrBr$_3$, VI$_3$, VSe$_2$, and Fe$_3$GeTe$_2$.~\citep{bedoyapinto2020intrinsic,Kim11131,acs.nanolett.9b00553,s41565-019-0565-0,s41567-019-0651-0, adma.201808074,s41565-018-0063-9,s41586-018-0626-9} Since the strength of magnetic anisotropies and the corresponding symmetry together dictate the underlying magnetic model, these 2D crystals with diverse anisotropy offer an opportunity to investigate the physics of different spin models, and test their applicability in real materials. Further, integration of these 2D ferromagnets in van der Waals heterostructures may trigger emergent phenomena, originating from the exchange proximity effects that could be utilized in disruptive technologies.~\citep{Zhonge1603113,Song1214,s41565-019-0629-1,PhysRevLett.124.197401,kezilebieke2020topological} 

Here, we particularly focus on the Cr-based ferromagnets, chromium trihalides and CrGeTe$_3$.~\citep{bedoyapinto2020intrinsic,nature22060,nature22391,Kim11131,acs.nanolett.9b00553,s41565-019-0565-0,s41567-019-0651-0} It was observed that the Curie temperature $T_{\C}$ increases with increasing perpendicular external magnetic field.~\citep{nature22060,nature22391} This phenomena, which is unprecedented in three-dimensional materials, indicates the significance of single-ion anisotropy in magnetic ordering on a 2D lattie. However, the magnitude of anisotropy is expected to be small from the minute Kerr rotation and less than 1 $\rm T$ spin reorientation field. This feature is also anticipated from the qualitative theoretical analysis. In all these materials, the Cr$^{3+}$ ions form in-plane honeycomb lattice and are subjected to octahedral crystal field. From Hund's rules, we expect $S=3/2$ state owing to $t_{2g}^3e_g^0$ electronic structure with a completely quenched orbital moment.~\citep{Kim11131,cm504242t,Carteaux_1995} Therefore, the single-ion anisotropy is expected to be very small below 1 meV, and only originate from the quantum fluctuation of the orbital moment in a distorted environment.~\citep{Lado_2017} The magnetic interactions between Cr$^{3+}$ ($S=3/2$) spins arise due to the superexchange interaction via the non-magnetic ligands proposed by Anderson.~\citep{PhysRev.79.350} According to the Goodenough-Kanamori rules,~\citep{GOODENOUGH1958287,KANAMORI195987} the magnetic ordering depends both on the filling of $d$-levels and the cation-anion-cation angle $\varphi$. Following this rules, ferromagnetic interaction emerges between the half-filled $t_{2g}^3$ spins, since $\varphi$(Cr$-X-$Cr) $\sim$ \ang{90}. Moreover, the corresponding theoretical description of $T_{\C}$ is strongly influenced by the presumed spin model, which is highly debated, and a complete picture is lacking.~\citep{Lado_2017,Torelli_2018, PhysRevB.98.144411,s41524-018-0115-6,PhysRevLett.124.017201,PhysRevB.101.134418, PhysRevMaterials.4.094004} Regardless of the expected inadequacy of single-ion anisotropy, CrI$_3$ was initially argued to have Ising behaviour.  The exact solution for the Ising model on the honeycomb lattice provides $k_{\rm B}T_{\C} = 1.519 J S^2$, where $J$ is the exchange interaction between the classical spins $S$. Thus, the Ising picture yields a considerably higher $T_{\C} $ of 90 - 130 K,~\citep{Lado_2017,Torelli_2018} which is much greater than the experimental value of 45 K.~\citep{nature22391} Strong renormalization in $T_{\C}$ is due to the spin fluctuations owing to inadequate single-ion anisotropy and thus, spins in CrI$_3$ should be treated within anisotropic Heisenberg model.~\citep{Lado_2017} A combination of various anisotropic interactions such as single-ion anisotropy, anisotropic exchange, Kitaev exchange and Dzialoshinskii-Moriya interaction has been considered with some  success.~\citep{Lado_2017, Torelli_2018, s41524-018-0115-6, PhysRevB.98.144411, PhysRevB.100.205409, PhysRevB.101.134418, PhysRevMaterials.4.094004,PhysRevB.102.115162, PhysRevLett.124.017201} Although, the magnetic interactions beyond the first-neighbour and their implications in 2D magnets are generally neglected in literature. 

Here, we develop a predictive yet straightforward spin model based on the first-principles calculations.  We show that the long-range magnetic exchanges beyond the first-neighbour along with the single-ion anisotropy provides a better description of magnetic ordering and also render the correct spin excitations in a honeycomb lattice. In the context of elusive room-temperature ferromagnetism, we predict a five-fold increase in $T_{\C}$ for the CrI$_3$ monolayer under experimentally attainable hole density. Further, the present model reproduces the experimentally observed trends in magnetism under carrier doping, and we provide a detailed microscopic mechanism. Moreover, we describe the effects of biaxial strain and chemical doping in the context of $T_{\C}$ manipulation. We also show that the present anisotropic Heisenberg model also characterizes the other similar materials such as CrCl$_3$, CrBr$_3$ and CrGeTe$_3$.

\section{Spin Hamiltonian}
To describe magnetism in these 2D materials, we propose a classical XYZ Heisenberg model Hamiltonian for the Cr$^{3+}$ ($S=3/2$) spins on a honeycomb lattice, 
\begin{equation}
\altmathcal{H} = -\frac{1}{2} \sum_{ij} \bigl[J_x S_i^xS_j^x  + J_y S_i^yS_j^y  + J_z S_i^zS_j^z \bigr] - \sum_i \altmathcal{A}_z S_i^zS_i^z. 
\nonumber
\end{equation}  
With $J_x = J_y = \altmathcal{J}$ as the isotropic exchange coupling and $\Lambda = J_z - \altmathcal{J}$ as the anisotropic exchange, the above Hamiltonian reduces to the XXZ Heisenberg Hamiltonian, 
\begin{equation}
\altmathcal{H} = -\frac{1}{2} \sum_{ij} \left(\altmathcal{J} \mathbf{S}_i \cdot \mathbf{S}_j  + \Lambda S_i^zS_j^z \right) - \sum_i \altmathcal{A}_z S_i^zS_i^z. 
\nonumber
\end{equation}

The isotropic exchange coupling $\altmathcal{J}$ between the neighbouring $\mathbf{S}_i$ and $\mathbf{S}_j$ spins favour either ferromagnetic $\altmathcal{J}>0$ or antiferromagnetic $\altmathcal{J}<0$ order. The symmetric term $\Lambda$ in the anisotropic superexchange interaction originates from the spin-orbit coupling of the anion. In contrast, the antisymmetric Dzyaloshinski-Moriya (DM) interaction, $-\altmathcal{D}_{ij} \cdot (\mathbf{S}_i \times \mathbf{S}_j)$ for the nearest-neighbor spins is zero due to the presence of inversion symmetry. Here, $\altmathcal{D}_{ij}$ is the DM vector. However, it has been shown that in a hexagonal lattice structure and in the presence of inversion symmetry, a finite intrinsic $\altmathcal{D}_{ij}$ perpendicular to the plane arises from the next-nearest-neighbor spin-orbit coupling. While such DM interaction may lead to interesting topological properties,~\citep{Owerre_2016,PhysRevLett.117.217202,PhysRevLett.117.227201,PhysRevX.8.041028} the next-nearest-neighbor DM interaction is at least an order of magnitude smaller than the magnetic anisotropies that are the lowest energy scales in the Hamiltonian.~\citep{PhysRevMaterials.4.094004} Therefore, it will be safe to ignore the DM interaction entirely in the present context of predicting magnetic phase transition. The $\altmathcal{A}_z$ is the single-ion anisotropy and calculated for the ferromagnetic ground state.  $\altmathcal{A}_z > 0$ and $\altmathcal{A}_z < 0$ indicate easy-axis and easy-plane magnetism, respectively. Since magnetic anisotropies are proportional to the spin-orbit coupling, we expect both $\Lambda$ and $\altmathcal{A}_z$ to be considerable in CrI$_3$ and CrGeTe$_3$, while these should be much smaller in CrBr$_3$ and CrCl$_3$. The other anisotropic terms such as the off-diagonal Kitaev interactions are not considered, since they are typically an order of magnitude smaller than the isotropic interaction.~\citep{s41524-018-0115-6}

\begin{figure}[!t]
\begin{center}
{\includegraphics[width=0.45\textwidth, angle=0]{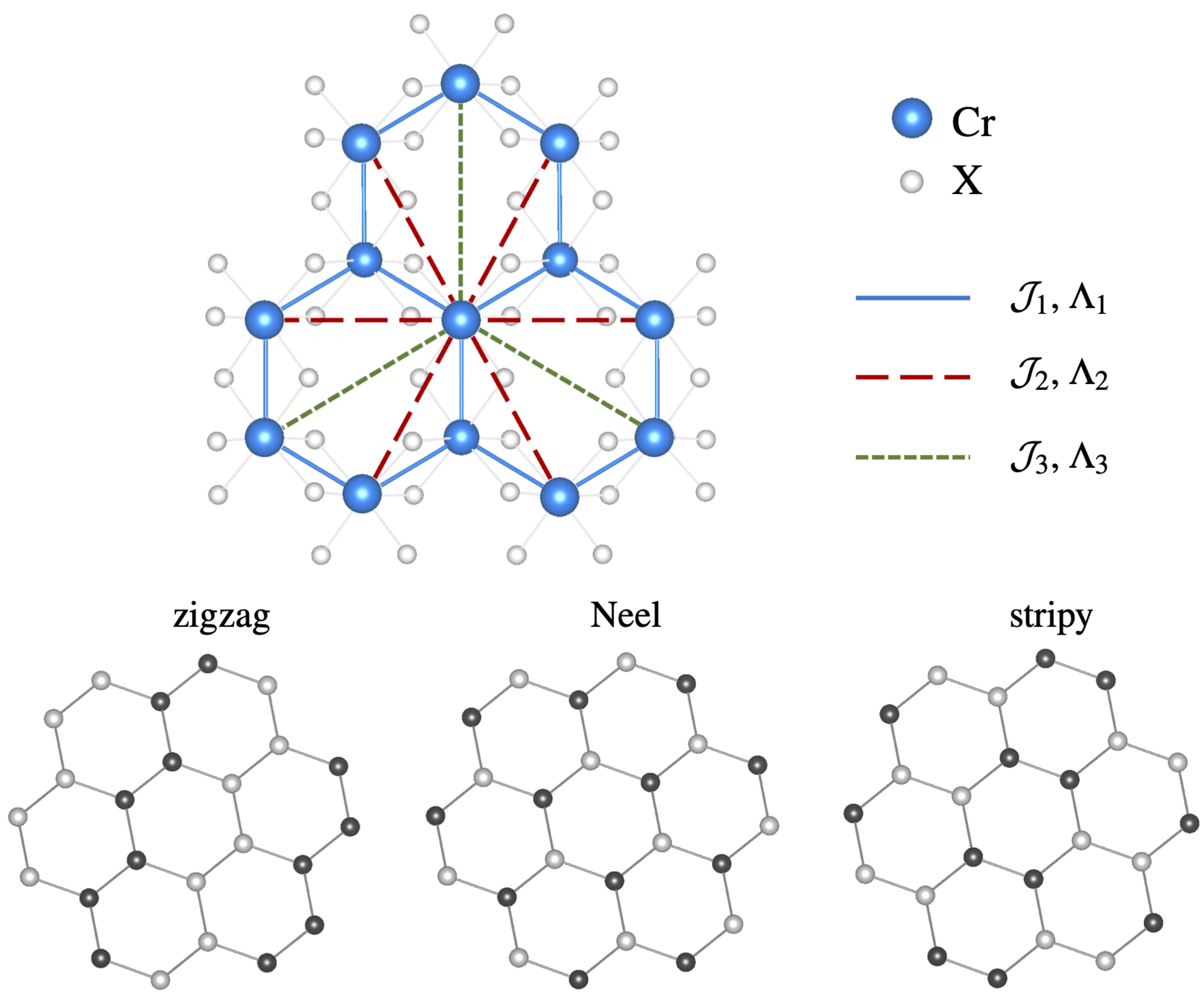}}
\caption{In the monolayers of Cr$X_3$ ($X=$ I, Br, Cl), the Cr$^{3+}$ ions are arranged in a honeycomb lattice of edge-sharing Cr$^{3+}$$X_6^{-}$ octahedra. The magnetic interactions $\altmathcal{J}_k$ and $\Lambda_k$ till the third-neighbour are marked, which are crucial to describe the magnetic ordering. The bottom panel schematically represents three different antiferromagnetic structures with zigzag, N\'{e}el and stripy spin orders in the honeycomb lattice. Dark and light-coloured balls characterise antiparallel Cr-spins.}
\label{fig:figure1}
\end{center}
\end{figure}

The above anisotropic Heisenberg Hamiltonian solely involving the nearest-neighbour exchange interactions have been used earlier.~\citep{Lado_2017, Torelli_2018, s41524-018-0115-6, PhysRevB.98.144411, PhysRevB.100.205409, PhysRevB.101.134418, PhysRevMaterials.4.094004,PhysRevB.102.115162, PhysRevLett.124.017201} While such a description provides a qualitative account, it typically underestimates the ordering temperature, and also fails to predict the correct experimental trends in magnetism under charge doping.\citep{Wang_2016} Moreover, such a nearest-neighbour consideration is not able to accommodate the complex spin excitations on the honeycomb lattice. These compel a spin model beyond the nearest-neighbour and consider the critical second- and third-neighbour interactions that can additionally describe the zigzag and stripy antiferromagnetic orders. Therefore, we extend the Hamiltonian to include the long-range isotropic and anisotropic exchange interactions, 

\begin{equation}
\altmathcal{H}_{\raisebox{-1.5pt}{\footnotesize$\ell$}} = -\frac{1}{2} \sum_{k=1}^3  \sum_{\langle ij \rangle_k} \bigl(\altmathcal{J}_k \mathbf{S}_{i} \cdot \mathbf{S}_{j}  + \Lambda_k S_{i}^zS_{j}^z \bigr) - \sum_i \altmathcal{A}_z S_i^zS_i^z,
\nonumber
\end{equation}

where $\langle ij \rangle_k$ with $k=1, 2, 3$ represents the first, second and third neighbours in the honeycomb magnetic lattice; $\altmathcal{J}_k$ and $\Lambda_k$ are the corresponding isotropic and anisotropic exchange interactions. Within the first-principles calculations, we will later show that that the long-range isotropic and anisotropic interactions are considerably large and comparable with the corresponding first-neighbour interactions. Further, these interactions become necessary to reproduce the correct experimental trend in the ordering temperature under carrier pumping.    

The parameters of the above spin Hamiltonian $\altmathcal{H}_{\raisebox{-1.5pt}{\footnotesize$\ell$}}$ are calculated within the density functional theory-based calculations as implemented in the Vienna {\em ab initio} simulation package.~\citep{PhysRevB.47.558,PhysRevB.54.11169}. Energies of the various magnetic configurations (Fig.~\ref{fig:figure1}) are fitted to the Heisenberg Hamiltonian $\altmathcal{H}_{\raisebox{-1.5pt}{\footnotesize$\ell$}}$. Additionally, different crystallographic orientations of the Cr-spins are also analyzed, by considering the relativistic spin-orbit coupling, to calculate various types of anisotropy energies. The wave functions within the spin-polarized density functional theory are described within the projector augmented formalism with the plane-wave basis of 600 eV for the kinetic energy cutoff.~\citep{PhysRevB.50.17953} The exchange-correlation energy is expressed by the local density approximation (LDA) that is supplemented with the Hubbard-like on-site Coulomb interaction $U$ for the localized Cr-3$d$ electrons. We choose $U_{\rm Cr}$ to be 0.5 eV for CrI$_3$ and CrGeTe$_3$ throughout the calculations using the rotationally invariant Dudarev's method.~\citep{PhysRevB.57.1505}  Relatively small value of $U_{\rm Cr}$ has been argued earlier,~\citep{s41524-018-0115-6} which results from the strong hybridization of Cr-$3d$ electrons with the shallow $p$-orbitals of the heavy ligands. In contrast, for CrBr$_3$ and CrCl$_3$, we use a slightly higher $U_{\rm Cr}$ of 1 and 1.5 eV, respectively. While the right set of exchange-correlation functional and $U_{\rm Cr}$ are required for better quantitative agreements with the experimental results, a different choice of exchange-correlation functional and $U_{\rm Cr}$ exhibit the same qualitative results. A 2$\times$2$\times$1 supercell with a vacuum space of 20 \AA\ perpendicular to the surface is adopted, which ensures negligible interactions between the periodic images. The structures are completely optimized until all the force components are below the $5\times10^{-3}$ eV/\AA\ threshold. The first Brillouin zone is sampled by a $\Gamma$ centred 17$\times$11$\times$1 Monkhorst-Pack $\mathbf{k}$-grid.~\citep{PhysRevB.13.5188}

To study the phase transition and obtain the $T_{\C}$, we implement Monte Carlo simulation on the 2D honeycomb lattice with the Hamiltonian $\altmathcal{H}_{\raisebox{-1.5pt}{\footnotesize$\ell$}}$. We consider a periodic 50 $\times$ 50 lattice consisting of $10^4$ spins that minimize the finite-size effects. The Metropolis algorithm is employed with a single-site update scheme.~\citep{10.1063/1.1699114} At every Monte Carlo step, a random spin $\mathbf{S}_i = (S^x, S^y, S^z)$  is selected, which is followed by an attempt to assign a new random direction in the three-dimensional space, $\mathbf{S}_i \rightarrow \mathbf{S}_i^{\prime}$, using the Marsaglia procedure.~\citep{marsaglia1972} At every temperature, we run 2 $\times$ 10$^8$ Monte Carlo steps to ensure thermal equilibrium, and 192 independent simulations are used to reduce the statistical fluctuation. The ordering temperature $T_{\C}$ is calculated by fitting the magnetic order parameter, $m(T) = m_0(1-T/T_{\C})^{\beta}$, where $\beta$ is the corresponding critical exponent.

\section{Results and Discussion}
First, we examine the magnetic ordering in chromium trihalides and CrGeTe$_3$ with honeycomb magnetic lattice, and extensively analyze the results in the context of the available experimental data. Once we establish the applicability of the present model to describe magnetism in these 2D materials, we will continue to discuss the electrical tunability of the corresponding magnetic ordering.

{\renewcommand{\arraystretch}{1.1}
\begin{table*}[!t]
\caption{Magnetic interaction parameters, in the anisotropic Heisenberg Hamiltonian $\altmathcal{H}_{\raisebox{-1.5pt}{\footnotesize$\ell$}}$, are obtained from the first-principles calculations including the spin-orbit coupling. The isotropic exchange $\altmathcal{J}_k$ beyond the first-neighbour has a crucial contribution. In the undoped systems, the single-ion and exchange anisotropies, $\altmathcal{A}_z$ and $\Lambda_k$, respectively, systematically increases with the spin-orbit coupling strength of the anion. The Curie temperatures $T_{\C}$ are calculated using the Monte Carlo simulations, and the results are in good agreement with available experimental data. The critical exponent $\beta$ deviates from the two-dimensional Ising model, and the values agree with the experimental predictions.}
\label{tab:table1}
\begin{tabular}{L{2.3cm}R{1.25cm}R{1.25cm}R{1.25cm}R{1.25cm}R{1.25cm}R{1.25cm}C{1.3cm}C{1.2cm}C{2.2cm}C{1.0cm}}
\hline
\hline
          & \multicolumn{3}{r}{Isotropic exchange  (meV)} & \multicolumn{3}{r}{Exchange anisotropy  ($\mu$eV)} & $\altmathcal{A}_z$ & \multicolumn{2}{c}{$T_{\C}$ (K)} & $\beta$ \\
          & $\altmathcal{J}_1$ & $\altmathcal{J}_2$ & $\altmathcal{J}_3$ & $\Lambda_1$ & $\Lambda_2$ & $\Lambda_3$ & (meV) & Present & Expt. &  \\ 
\hline
CrI$_3$ &	2.358 &	0.640 & $-$0.636 &	$-$55.6 & $-$55.6	& 155.5	& 0.289	& 42.5	& 45~\citep{nature22391}	& 0.215 \\
CrBr$_3$ &	1.933 &	0.361 & $-$0.322 &	9.7 & $-$5.0	& 21.2	& 0.082	& 30.5	& 27-34~\citep{Kim11131,acs.nanolett.9b00553}	& 0.226 \\
CrCl$_3$      & 1.142  & 0.165      & $-$0.234& 0 & 0 & 3.7 & 0.007& 13 & 13~~\citep{bedoyapinto2020intrinsic} & 0.287 \\
CrGeTe$_3$ & 5.522  & $-$0.642 & 0.356 & 112.2 & 19.4 & 81.5 & 0.178 & 53  & 30 $\pm$ 5 (2L)~\citep{nature22060}& 0.213 \\
Mo$_{0.25}$Cr$_{0.75}$I$_3$ & 2.392  & 0.551 & $-$0.866 & $-$38.9 & $-$63.9 & 135.2 & 0.433 & 33 & ---  &  0.182 \\
W$_{0.25}$Cr$_{0.75}$I$_3$  & 3.186  & 0.596 & $-$0.731 & $-$38.9 & $-$75.0 & 101.9 & 0.778 & 50.5 & ---  & 0.205 \\
\hline
\hline
\end{tabular}
\end{table*}}

\subsection{Magnetic ordering in two-dimension}
Calculations indicate that CrI$_3$ is a ferromagnetic insulator, and the magnetic moment is localized mostly at the Cr-sites on the honeycomb lattice. Various antiferromagnetic excited states persist within 7-16 meV/CrI$_3$ energy, and the zigzag antiferromagnetic order is found to be the first excited state independent of exchange-correlation functional and on-site Coulomb interaction (Supplemental Material~\citep{supple}). Such zigzag ground state is found in honeycomb antiferromagnets such as FePS$_3$,~\citep{nanolett.6b03052} NiPS$_3$,~\citep{PhysRevB.92.224408,s41467-018-08284-6} and quantum spin liquid candidates RuCl$_3$~\citep{PhysRevB.92.235119} and Na$_2$IrO$_3$.~\citep{PhysRevLett.110.097204} In contrast, the previous calculations do not consider such spin excitation and the exchange interactions were extracted by considering the excited N\'{e}el order.~\citep{Lado_2017,Torelli_2018} The Cr-spins preferentially align perpendicular to the lattice plane, and the corresponding single-ion anisotropy $\altmathcal{A}_z$ is 0.29 meV indicating easy-axis anisotropy (Table~\ref{tab:table1}), which is in agreement with experiments.~\citep{nature22391} Such a small $\altmathcal{A}_z$ is indicative of quenched orbital moment at Cr with $t_{2g}^3e_g^0$ electronic configuration and is consistent with the previous calculations.~\citep{Lado_2017,Torelli_2018,s41524-018-0115-6} Moreover, the calculated $\altmathcal{A}_z$ is in good quantitative agreement with the experimental estimate of 0.27 meV from the acoustic magnon energy shift interpolated to zero field.~\citep{s41567-020-0999-1} The isotropic and anisotropic exchange interactions, $\altmathcal{J}_k$ and $\Lambda_k$, are calculated by mapping the energies of the ferromagnetic ground state along with the zigzag, N\'{e}el and stripy antiferromagnetic orders with the anisotropic Heisenberg Hamiltonian $\altmathcal{H}_{\raisebox{-1.5pt}{\footnotesize$\ell$}}$ (Table~\ref{tab:table1}). Both the in-plane and out-of-plane spin orientations are explicitly considered in this regard.

The exchange interactions beyond the first-neighbour are substantial, $|\frac{\altmathcal{J}_2}{\altmathcal{J}_1}| = $ 0.27 and $|\frac{\altmathcal{J}_3}{\altmathcal{J}_2}| = $ 1, for CrI$_3$ that can not be neglected (Table~\ref{tab:table1}). Moreover, while the first- and second-neighbour interactions are ferromagnetic, $\altmathcal{J}_3$ is antiferromagnetic. The anisotropic exchange interactions $\Lambda_kS_i^zS_j^z$ appear due to the strong spin-orbit coupling at the halide site. With these first-principles parameters $\altmathcal{J}_k$, $\Lambda_k$, and $\altmathcal{A}_z$ in the anisotropic XXZ Heisenberg Hamiltonian $\altmathcal{H}_{\raisebox{-1.5pt}{\footnotesize$\ell$}}$ (Table~\ref{tab:table1}), we calculate the $T_{\C}$ using the Metropolis Monte Carlo method. We predict that the ferromagnetic ordering emerges at 42.5 K, which is in excellent agreement with the experimental $T_{\C}$ of 45 K.~\citep{nature22391} It would be interesting to further decipher the quantitative importance of the second- and third-neighbour interactions in $\altmathcal{H}_{\raisebox{-1.5pt}{\footnotesize$\ell$}}$. We notice that the ordering temperature is severely underestimated to 28 K, while only the first-neighbour interactions, $\altmathcal{J}_1\mbox{-}\Lambda_1\mbox{-}\altmathcal{A}_z$, are considered. The inclusion of ferromagnetic $\altmathcal{J}_2$ increases the ordering temperature and the corresponding $\altmathcal{J}_1\mbox{-}\Lambda_1\mbox{-}\altmathcal{J}_2\mbox{-}\Lambda_2\mbox{-}\altmathcal{A}_z$ model overestimates the  $T_{\C}$ to 55 K.
Therefore, for a better quantitative description of magnetic ordering, the inclusion of interactions till the third-neighbour is essential. However, the long-range exchange interactions are largely neglected in the context of 2D magnetism.~\citep{Lado_2017, Torelli_2018, s41524-018-0115-6, PhysRevB.98.144411, PhysRevB.100.205409, PhysRevB.101.134418, PhysRevMaterials.4.094004,PhysRevB.102.115162, PhysRevLett.124.017201}  To understand the critical behaviour of the system, we calculate $\beta$ by fitting $m(T) = (1-T/T_{\C})^{\beta}$ power-law near the phase transition and we estimate $\beta$ to be 0.215 $\pm$ 0.002. While there is no experimental estimation for the monolayer, consistent with the expected $\beta_{\rm 2D} < \beta_{\rm 3D}$ feature, the present result is slightly lower than the bulk CrI$_3$, $\beta_{\rm 3D} =$ 0.249 $\pm$ 0.014 estimated via inelastic neutron scattering experiments.~\citep{PhysRevB.97.014420,PhysRevB.101.134418}

\begin{figure}[!t]
\begin{center}
{\includegraphics[width=0.39\textwidth, angle=0]{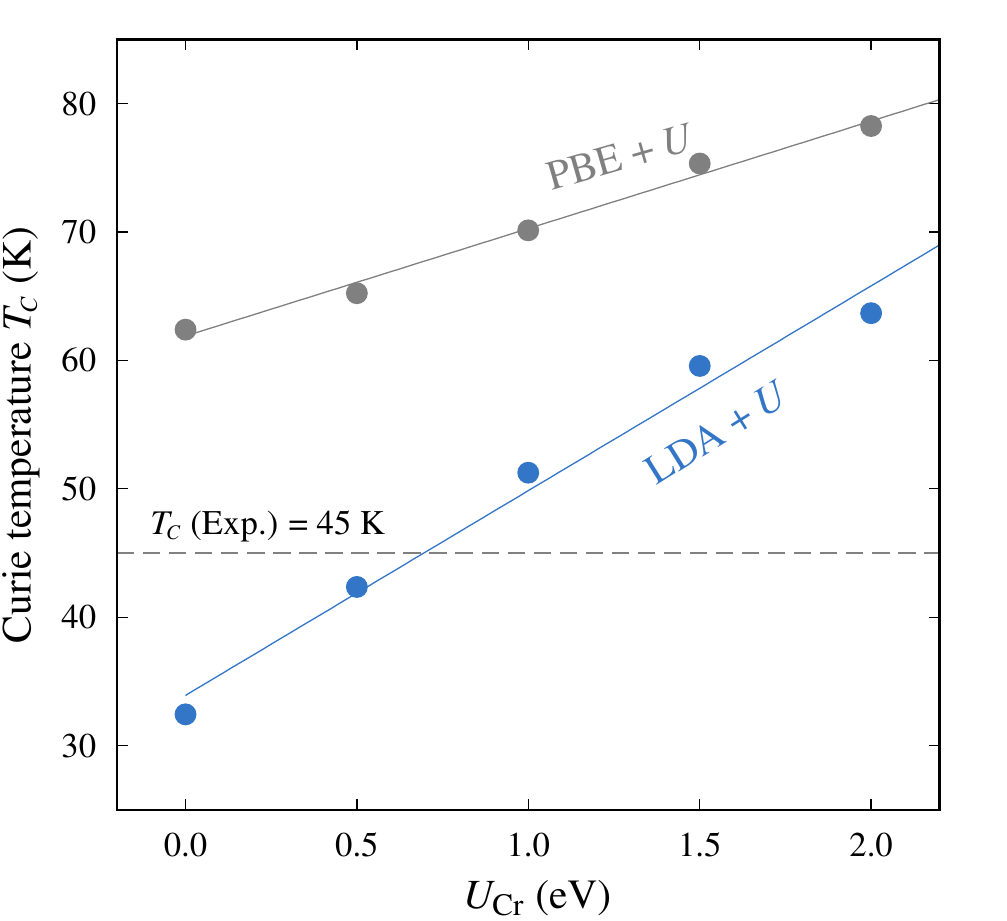}}
\caption{Magnetic interactions and therefore the $T_{\C}$, crucially depend on the type of exchange-correlation functional and the on-site Coulomb interaction $U_{\rm Cr}$.
Magnetism in CrI$_3$ is better described by the LDA + $U$ formalism. The calculated $T_{\C}$ of 42.5 K with the first-principles parameters and $U_{\rm Cr}=$ 0.5 eV is in excellent agreement with the experiment.~\citep{nature22391} The spin-orbit coupling is considered for all the calculations.     
}
\label{fig:figure2}
\end{center}
\end{figure}

We observe that while a different set of exchange-correlation functional and on-site Coulomb interaction provides a similar qualitative result, the ordering temperature is crucially dependent on the choice of $U_{\rm Cr}$ (Fig.~\ref{fig:figure2}).  In this regard, we investigate CrI$_3$ with LDA and Perdew-Burke-Ernzerhof (PBE) exchange-correlation functionals and varied the on-site Hubbard interaction within 0$-$2 eV range.~\citep{supple} Independent of the functional and $U_{\rm Cr}$, the qualitative results such as the ferromagnetic ground state and the signs of $\altmathcal{J}_k, \Lambda_k$, and $\altmathcal{A}_z$, remain unaltered. However, their magnitudes are greatly affected. Regardless the exchange-correlation description, the ferromagnetic $\altmathcal{J}_1$ increases monotonically with $U_{\rm Cr}$, $\altmathcal{J}_2$ remains unaffected, and the magnitude of antiferromagnetic $\altmathcal{J}_3$ decreases slowly.~\citep{supple} In contrast, the anisotropic interactions are not significantly affected by $U_{\rm Cr}$. Combined variations in the interaction parameters are reflected in the calculated $T_{\C}$ that increases with the on-site Hubbard interaction (Fig.~\ref{fig:figure2}). Further, we notice that the PBE functional systematically overestimates the $T_{\C}$ due to a much stronger first-neighbour isotropic $\altmathcal{J}_1$.  

The present results can be discussed in comparison with the Ising model. Note that, in the limit of $\altmathcal{A}_z \rightarrow \infty$, the in-plane spin components are quenched, and all excitations will have spins aligned along the out-of-plane easy-axis. Consequently, the present spin model in $\altmathcal{H}_{\raisebox{-1.5pt}{\footnotesize$\ell$}}$ becomes asymptotically equivalent to the Ising model with $J_k^{\rm Ising} = \altmathcal{J}_k + \Lambda_k$. Assuming the first-neighbour interaction only, we estimate $T_{\C}^{\rm Ising} = 1.519(\altmathcal{J}_1 + \Lambda_1) \sim$ 91 K, which is also consistent with the Ising Monte Carlo simulations.  However, $T_{\C}^{\rm Ising}$ is much higher than the experimental ordering temperature of 45 K, and the present prediction of 42.5 K within the anisotropic Heisenberg model. Further inclusion of the long-range interactions aggravates the situation, and through the Monte Carlo simulations, we obtain $T_{\C}^{\rm Ising}$ as 132 K, within the $J_1 \mbox{-} J_2 \mbox{-} J_3$ Ising model. In contrast, the rapid spin fluctuation owing to the small $\altmathcal{A}_z$ is responsible for the strong renormalization of ordering temperature in CrI$_3$.

Beyond the classical approximation, the Cr-spins can be treated by quantum mechanical $S=3/2$ operators. Within the linear spin-wave approximation, the spin operators are expressed in terms of Holstein-Primakoff transformation.~\citep{PhysRev.58.1098} At low temperature $S^z \simeq S$, and the spin excitations become gapped due to the magnetic anisotropy that is responsible for magnetic ordering in 2D. Within this formalism, the zero-temperature spin wave gap can be expressed as,~\citep{Lado_2017,Torelli_2018} $\Delta_0 = (2S-1) \altmathcal{A}_z + \sum_k N_k S \Lambda_k$, where $N_k$ is the number of $k$-th nearest-neighbours; $N_k=$ 3, 6, 3 for the first, second and third nearest-neighbours, respectively, for the honeycomb lattice. The calculated spin gap of $\Delta_0=$ 0.54 meV is in good agreement with the most recent experimental data such as the ferromagnetic resonance (0.3 meV),~\citep{PhysRevLett.124.017201} inelastic neutron scattering experiment (0.37 meV),~\citep{PhysRevB.101.134418} and magneto-Raman spectroscopy.~\citep{s41567-020-0999-1}

Having discussed the case of CrI$_3$ in details, it would be straightforward to describe the magnetic phase transitions in the other similar materials of interest, CrBr$_3$, CrCl$_3$ and CrGeTe$_3$. While CrBr$_3$ and CrCl$_3$ are isostructural to CrI$_3$, the magnetic lattice of CrGeTe$_3$ also has honeycomb symmetry. Similar to CrI$_3$, in all these  materials, the ferromagnetic ordering emerges between the Cr$^{3+} (S=3/2)$ spins according to the Anderson-Goodenough-Kanamori rules. The corresponding ordering temperature depends on their respective interaction parameters in the Hamiltonian $\altmathcal{H}_{\raisebox{-1.5pt}{\footnotesize$\ell$}}$. While Cr-ions in these materials have the same $t_{2g}^3e_{g}^0$ electronic configuration, due to a reduction in the spin-orbit coupling strength at the halide side along I $\rightarrow$ Br $\rightarrow$ Cl row, the magnitudes of both $\altmathcal{A}_z$ and $\Lambda_k$ are expected to decrease accordingly (Table~\ref{tab:table1}), indicating a reduced preference to the out-of-plane orientation of the Cr moments. This trend has been experimentally demonstrated in mixed-halide CrCl$_{3-x}$Br$_x$ compounds.~\citep{adma.201801325} The present results that includes the relativistic interactions are in  qualitative agreement with earlier theoretical efforts, although different exchange-correlation functionals were used.~\citep{PhysRevB.98.144411,PhysRevB.100.205409,PhysRevB.102.115162}   

Compared to CrI$_3$, the lowering of spin-orbit coupling at the halide site results in smaller anisotropic magnetic interaction parameters $\Lambda$ and $\altmathcal{A}_z$ for CrBr$_3$ (Table~\ref{tab:table1}). However, the nature of the isotropic interactions $\altmathcal{J}_k$ remain unchanged. The Monte Carlo simulation reveal a ferromagnetic ordering temperature of 30 K, which is in good agreement with the experimental estimates  of 27-34 K determined using the magneto-optical~\citep{Kim11131,acs.nanolett.9b00553} and micro-magnetometry measurements.~\citep{s41928-019-0302-6} Since the bulk CrBr$_3$ orders at 37 K,~\citep{JPSJ.15.1664} the relatively small decrease in $T_{\C}$ in monolayer suggest that the interlayer interaction in bulk and few-layer flakes is weak. The magnetization $m(T)$ follows $(1-T/T_{\C})^{\beta}$ power-law and the best fit near the $T_{\C}$ yields $\beta=$0.226, which is consistent with the monolayer CrI$_3$ (Table~\ref{tab:table1}).

While layered antiferromagnetism persists till the bilayer CrCl$_3$,~\citep{Kim11131,s41565-019-0565-0} the magnetism in monolayer remained elusive until recently.~\citep{bedoyapinto2020intrinsic} The ordering temperature is recognised to be almost thickness independent,~\citep{Kim11131,s41565-019-0565-0} and CrCl$_3$ has been regarded as a two-dimensional easy-plane spin system with weak anisotropy, which is in contrast to both CrI$_3$ and CrBr$_3$. However, theoretical calculations, including the earlier attempts,~\citep{PhysRevB.98.144411,PhysRevB.100.205409,PhysRevB.102.115162} are in contradiction, and we predict a weak easy-axis anisotropy (Table~\ref{tab:table1}). It was argued that magnetic anisotropy is comparable to the magnetic shape anisotropy in CrCl$_3$, which ultimately dictates the overall anisotropy to be in-plane.~\citep{PhysRevB.100.224429} Moreover, the present results on the free-standing monolayer may not be immediately compared to the recent experiment since CrCl$_3$ was deposited on graphene/6H-SiC(0001), resulting in substrate-induced trigonal distortions.~\citep{bedoyapinto2020intrinsic} Such distortion may induce a stronger magnetic anisotropy compared to the present calculation. The nature of $\altmathcal{J}_k$ follows similar trend, and second and third neighbour interactions are still significant (Table~\ref{tab:table1}). However, the exchange anisotropies are found to be almost zero due to very weak halide spin-orbit coupling. Regardless the contradiction in $\altmathcal{A}_z$, the calculated Cr-moment of 2.87 $\mu_B$, $T_{\C}$ and $\beta$ are in good agreement with the experimental data.~\citep{bedoyapinto2020intrinsic} Using the magnetic interaction parameters in Table~\ref{tab:table1}, we estimate the ferromagnetic $T_{\C}$ to be 13 K, which is in good agreement with 13 K determined by the element-specific X-ray magnetic dichroism.~\citep{bedoyapinto2020intrinsic} Near the magnetic transition, the exponent $\beta$ (0.287 $\pm$ 0.021) in CrCl$_3$ is larger than CrI$_3$ and CrBr$_3$ monolayers.


The case of CrGeTe$_3$ is interesting, which in bulk orders around 65 K~\citep{PhysRevB.95.245212,PhysRevB.98.214420,PhysRevB.100.134437,s41565-018-0186-z} that is slightly higher than the CrI$_3$ counterpart.~\citep{cm504242t} While a layer-dependent $T_{\C}$ is demonstrated, experimental observation of ordering has remained absent in the monolayer.~\citep{nature22060} We notice that the monolayer CrGeTe$_3$ has easy-axis anisotropy in agreement with the bulk and few-layer samples.~\citep{PhysRevB.95.245212,PhysRevB.98.214420,PhysRevB.100.134437,s41565-018-0186-z} Expectedly the $\altmathcal{A}_z$ is weaker than that of CrI$_3$ (Table~\ref{tab:table1}). The nature of exchange coupling is slightly different from the chromium trihalides; $\altmathcal{J}_1$ remains ferromagnetic but with $\altmathcal{J}_2 < 0$ and $\altmathcal{J}_3 > 0$.  The magnetic parameters from the first-principles calculations (Table~\ref{tab:table1}) indicate expected ordering, which contrasts the experimental data to date.  Within the anisotropic Heisenberg Monte Carlo, we determine the $T_{\C}$ to be 53 K, which is somewhat higher than the bilayer (30 $\pm$ 5 K) but comparable to the few-layer flakes, 52 $\pm$ 5 K for 5-layer CrGeTe$_3$.~\citep{nature22060} In the bulk crystal, the Cr-moment decreases rapidly with increasing temperature, from 2.8 $\mu_B$ (2 K) to 2.3 $\mu_B$ at 60 K.~\citep{PhysRevB.100.134437}  Considering this fact, we have recalculated the $T_{\C}$ using a smaller Cr-moment of 2.3 $\mu_B$ along with the exchange parameters in the Table~\ref{tab:table1}, which produces a lower $T_{\C}$ of 30.5 K ($\beta$ = 0.191), which is in better agreement with the experiment. The calculated $\beta$ for the monolayer is 0.213, which is in agreement with the bulk values of 0.177-0.242 that are experimentally determined.~\citep{PhysRevB.95.245212,PhysRevB.98.214420} While we anticipate a ferromagnetic ordering in the monolayer according to the present results and also expect the same from the experimental trends in few-layer flakes,~\citep{nature22060} it has remained practically elusive till date. Moreover, we will discuss later that accidental chemical contamination or passivation leading to small charge doping may destroy ordering in the monolayer limit. Therefore, in this regard, monolayer CrGeTe$_3$ should draw further theoretical and experimental attention.

\begin{figure}[!t]
\begin{center}
{\includegraphics[width=0.48\textwidth, angle=0]{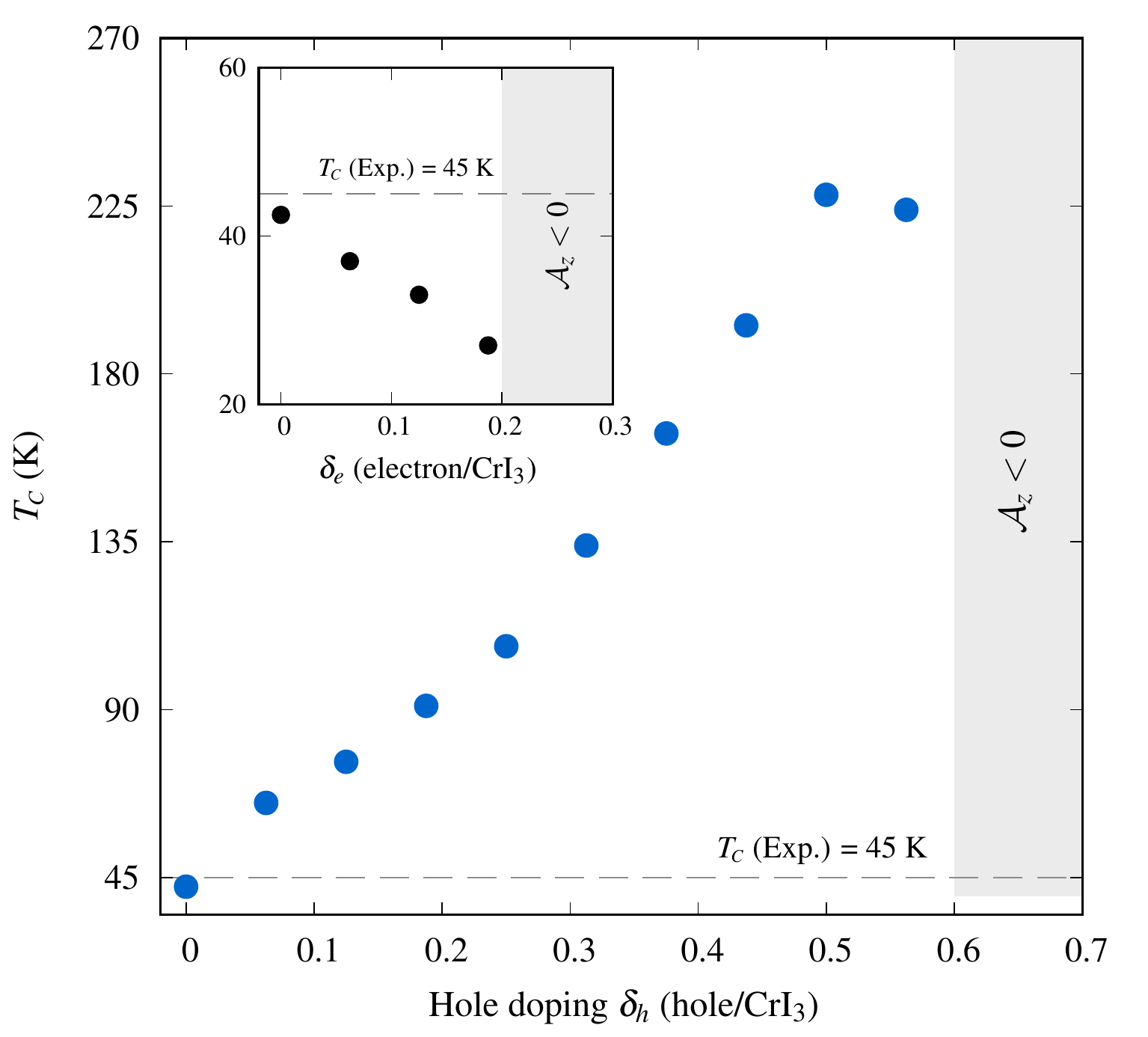}}
\caption{The Curie temperature $T_{\C}$ of single-layer CrI$_3$ is critically affected by the carrier doping, which could be electrically exploited. The experimental $T_{\C}$ of 45 K for the neutral monolayer CrI$_3$~\citep{nature22391} is well reproduced within the present long-range and anisotropic Heisenberg model. The hole and electron (inset) doping has contrasting effects. Calculated $T_{\C}$ gradually increases with hole doping, and the increase is as much as five-fold  for $\delta_h$ = 0.5 hole/CrI$_3$, which is equivalent to 2.66 $\times$ 10$^{14}$ cm$^{-2}$ hole density. Such a level of doping could be achieved through ionic gating or femtosecond laser pulse. In contrast, the electron doping, shown in the inset, has detrimental effect, since the $T_{\C}$ gradually decreases with increasing doping level $\delta_e$. The overall trends are in agreement with the experimental observations.~\citep{s41565-018-0135-x} Beyond certain doping densities, the Cr-moments prefer in-plane orientations with $\altmathcal{A}_z < 0$.
}
\label{fig:figure3}
\end{center}
\end{figure}

\begin{figure*}[!t]
\begin{center}
{\includegraphics[width=0.995\textwidth, angle=0]{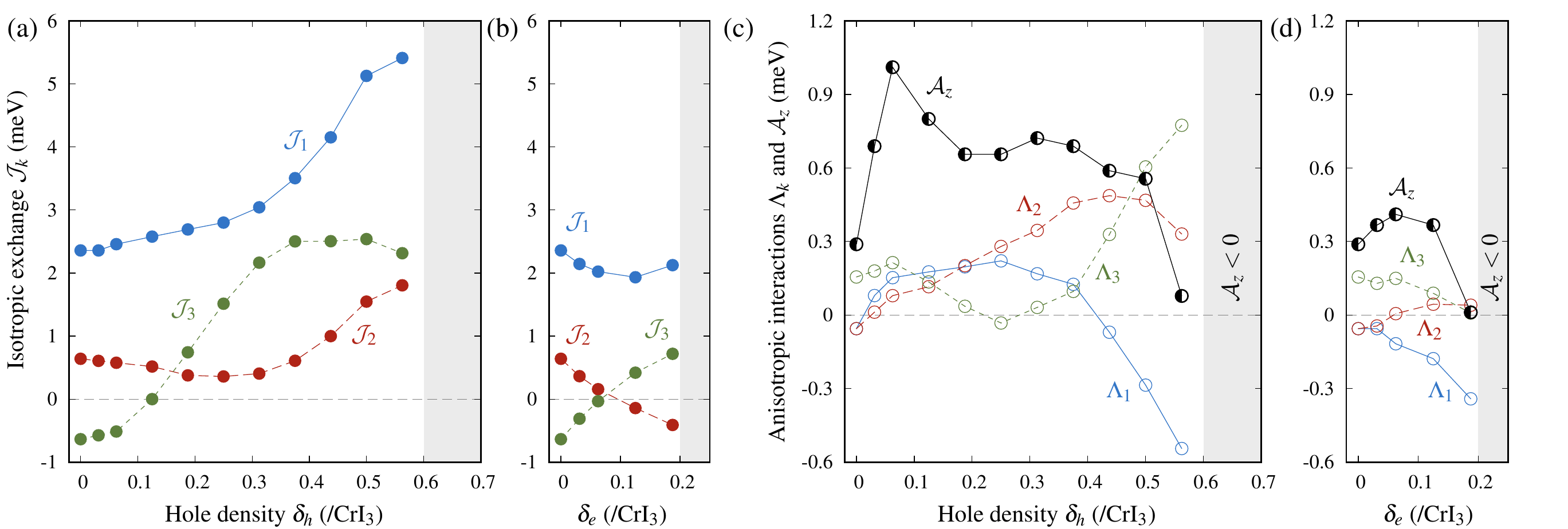}}
\caption{For the monolayer CrI$_3$, the microscopic dependence of the various magnetic interactions on carrier doping that can be electrically controlled. (a) and (b) show the variations in the isotropic exchange interactions $\altmathcal{J}_k$ with hole and electron doping, respectively. $\altmathcal{J}_k$ are severely affected by carrier doping. (c) and (d) represent the variations in the anisotropic magnetic interactions $\Lambda_k$ and $\altmathcal{A}_z$ with $\delta_h$ and $\delta_e$, respectively, which are mostly non-monotonous. The complex interdependence of orbital occupancies, the local and orbital moments, and the resulting on-site spin-orbit coupling drive the intricate doping dependence. Beyond the critical hole and electron density, the magnetism switches, and the Cr-moments prefers in-plane orientation, $\altmathcal{A}_z < 0$, which is represented by the shaded region. These variations in  $\altmathcal{J}_k$, $\Lambda_k$ and $\altmathcal{A}_z$ dictate the discussed doping-dependent ordering temperature $T_{\C}$ (Fig.~\ref{fig:figure3}), and qualitatively corroborates the experimental observations.~\citep{s41565-018-0135-x}
}
\label{fig:figure4}
\end{center}
\end{figure*}

\subsection{Manipulation of magnetism}
Electric control of magnetism is of fundamental interest in nanoscale magnetic devices.~\citep{nnano.2015.22}  Interlayer magnetic coupling in bilayer CrI$_3$ and the strength of magnetic order in three-layer Fe$_3$GeTe$_2$ have been altered by inducing charge via electrostatic gating.~\citep{s41565-018-0121-3,s41565-018-0135-x,s41586-018-0626-9} A three-fold increase in the ferromagnetic ordering temperature is reported in three-layer itinerant Fe$_3$GeTe$_2$ flakes.  Therefore, it would be worthwhile to explore the effects of electrostatic gating in these 2D materials.  

We observe contrasting consequences of hole and electron doping on the ordering temperature (Fig.~\ref{fig:figure3}), and the present results also corroborate the experimental observations under electrostatic gating. Hole doping strengthens the ferromagnetic order in monolayer CrI$_3$, while electron doping weakens magnetism (Fig.~\ref{fig:figure3}). Calculated Curie temperature $T_{\C}$ increases monotonically upon hole injection, and rises as much as five-fold to 228 K  at an experimentally achievable hole density of $\delta_h$ = 0.5 hole/CrI$_3$ $\sim$ 2.66 $\times$ 10$^{14}$ cm$^{-2}$. In contrast, electron doping shows deleterious influence, as  $T_{\C}$ consistently decreases to 27 K at a density $\delta_e$ = 0.2 electron/CrI$_3$ (Fig.~\ref{fig:figure3}). We also observe that at a critical electron and hole densities the magnetism in monolayer CrI$_3$ switches to in-plane orientation, $\altmathcal{A}_z < 0$. The present results correspond well with the experimental trends, where differing impacts were discerned upon electron and hole doping.~\citep{s41565-018-0135-x} At a low doping level of about $\delta \sim$ 2.5 $\times$ 10$^{13}$ cm$^{-2}$, a 10\% rise or reduction in $T_{\C}$ was realised via hole and electron doping, respectively.~\citep{s41565-018-0135-x}  While a charge density below 3 $\times$ 10$^{13}$ cm$^{-2}$ has been experimentally realised,~\citep{s41565-018-0135-x} a much higher density beyond 10$^{14}$ cm$^{-2}$ should be feasible through ionic gating or femtosecond laser pulse in CrI$_3$. For example, very high electrostatic doping of 4 $\times$ 10$^{14}$ cm$^{-2}$ has been obtained in monolayer MoS$_2$ via ionic liquid gating.~\citep{ncomms9826,nnano.2015.314} Further, a very high density of carriers $\sim$ 3.2 $\times$ 10$^{14}$ cm$^{-2}$ were injected via femtosecond pulse laser in a few-layer ferromagnetic Fe$_3$GeTe$_3$.~\cite{PhysRevLett.125.267205} 

The present trends described in Fig.~\ref{fig:figure3}  are in contradiction with the earlier theoretical predictions, where an increase in ordering temperature was perceived in both electron and hole-doped monolayers.~\cite{Wang_2016} We argue this disagreement to correlate with an inadequate description of magnetism considered earlier.  In this context, it is intriguing to understand the situation if we only considere the first-neighbour interaction, and calculate $\altmathcal{J}_1$ and $\Lambda_1$ using the energy difference, $\Delta_{FM}^N = E_{\mathrm {Neel}} - E_{\mathrm {FM}}$ (Supplemental Material~\citep{supple}); similar to that in Ref.~\cite{Wang_2016}.  Qualitative analysis indicates that under such incomplete magnetic description, both electron and hole doping contributes to the increase in ordering temperature. However, the minimalistic picture of only first-neighbour interaction fails to reproduce the correct experimental dependance,~\citep{s41565-018-0135-x} along with the present results (Fig.~\ref{fig:figure3}).
This analysis further establishes the importance of long-range exchange interactions in predicting the ordering temperature under electrostatic doping.   

Although an observation similar to that described in Fig.~\ref{fig:figure3} has been made experimentally, the microscopic mechanism of doping dependent $T_{\C}$ remains unclear.~\citep{s41565-018-0135-x}  The present results reveal that the fundamental interdependence of various magnetic interactions with doping (Fig.~\ref{fig:figure4}) is responsible for the observed trend. We find that charge doping severely alters the isotropic exchange couplings $\altmathcal{J}_k$ as well as both types of anisotropic interactions $\Lambda_k$ and $\altmathcal{A}_z$ (Fig.~\ref{fig:figure4}). The isotropic $\altmathcal{J}_1$ and $\altmathcal{J}_2$ remains ferromagnetic upon hole doping and follow a similar behaviour [Fig.~\ref{fig:figure4}(a)]. Below a critical hole density of $\delta_h < 0.4$ hole/CrI$_3$, $\altmathcal{J}_1$ and $\altmathcal{J}_2$ are only slightly affected, and beyond this density these exchange interactions increase very rapidly. In contrast, the antiferromagnetic $\altmathcal{J}_3$ is critically affected by the hole doping with $\delta_h < 0.4$ hole/CrI$_3$, but remains unaltered with further increase in doping level [Fig.~\ref{fig:figure4}(a)].  At low doping, the antiferromagnetic $\altmathcal{J}_3$ monotonically increases with hole density and changes sign to become ferromagnetic.  In the case of electron doping, $\altmathcal{J}_1$ decreases with increasing doping, while $\altmathcal{J}_2$ and $\altmathcal{J}_3$ show opposite trends [Fig.~\ref{fig:figure4}(b)]. The isotropic $\altmathcal{J}_2$ ($\altmathcal{J}_3$) decreases (increases) monotonically with $\delta_e$ and consequently changes the nature of magnetic coupling at a particular density.  

Charge doping perturbs the orbital occupancies that are reflected in the calculated moments localized at the Cr-sites. The $|\mu_{\mathrm Cr}|$ decreases gradually with hole doping, and the trend is reversed for the electron doping. However, the change is only about $\pm$5\% within the entire doping range considered in Fig.~\ref{fig:figure3}.  In analogy with an earlier experiment, a similar trend is witnessed in the saturation magnetization.~\citep{s41565-018-0135-x} Compared to $\mu_{\mathrm Cr}$, the induced $|\mu_{\mathrm I}|$ moment is severely affected by the charge doping but shows a trend opposite to the $\mu_{\mathrm Cr}$.  The $|\mu_{\mathrm I}|$  is 0.04 $\mu_{B}$  for the neutral case, which increases to 0.06 $\mu_{B}$  at 0.5 hole/CrI$_3$, and decreases to 0.03 at 0.25 electron/CrI$_3$. The orbital moments $\mu_{\rm orb}^{\rm Cr}$ and $\mu_{\rm orb}^{\rm I}$ are also affected by the charge doping. The $\mu_{\rm orb}^{\rm Cr}$ ($\mu_{\rm orb}^{\rm I}$) decreases (increases) to 0.047 $\mu_B$ (0.017 $\mu_B$) at $\delta_h=$ 0.5 hole/CrI$_3$ from 0.091 $\mu_B$ (0.008 $\mu_B$) for the neutral CrI$_3$ monolayer. In contrast, the orbital moments are not affected significantly upon electron doping.

Therefore, orbital occupancies, local and orbital moments together perturb the on-site spin-orbit coupling energies that eventually trigger modifications in various anisotropic magnetic interactions [Fig.~\ref{fig:figure4}(c) and \ref{fig:figure4}(d)]. Overall, the variations in $\Lambda_k$ and $\altmathcal{A}_z$ are non-monotonous. The anisotropic exchange interactions $\Lambda_1$ and $\Lambda_2$ are negative for the neutral CrI$_3$, which change sign with hole doping and remain positive except $\Lambda_1 < 0$ beyond 0.4 hole/CrI$_3$ [Fig.~\ref{fig:figure4}(c)]. In contrast, $\Lambda_3$ remains positive and raises sharply for $\delta_h > 0.4$  hole/CrI$_3$. The single-ion anisotropy $\altmathcal{A}_z$ increases sharply with hole doping at low density, which then gradually falls with a further increase in $\delta_h$. Finally, the nature of magnetism changes completely at very high hole density $\delta_h > 0.6$  hole/CrI$_3$ and the Cr-spins prefer in-plane orientation with $\altmathcal{A}_z < 0$ [Fig.~\ref{fig:figure4}(c)].  Electron doping, on the other hand, has contrasting behaviour.  While $\Lambda_1$ and $\Lambda_3$ decreases with increasing doping, the $\Lambda_2$ increases in comparison and changes sign at a density [Fig.~\ref{fig:figure4}(d)]. Compared to the hole doping, $\altmathcal{A}_z$ shows a similar trend due to electron doping. However, the XY-type spin preference is observed at a much lower density $\delta_e$ [Fig.~\ref{fig:figure3}(d)]. The combinatorial doping dependencies of $\altmathcal{J}_k$, $\Lambda_k$ and $\altmathcal{A}_z$ result in contrasting doping-dependant $T_{\C}$ for the hole and electron doping (Fig.~\ref{fig:figure3}), and also corroborate the microscopic mechanisms behind the similar experimental observations.~\citep{s41565-018-0135-x} 

The multiorbital picture of superexchange interactions can also corroborate the doping-dependent $T_{\C}$ in Fig.~\ref{fig:figure3}. The Cr$^{3+}$-ions with $t_{2g}^3e_g^0$ electronic configurations interact via $t_{2g}^3-t_{2g}^3$ and $t_{2g}^3-e_{g}^0$ orbital channels. According to the Kugel-Khomskii mechanism, the half-filled $t_{2g}^3-t_{2g}^3$ exchange interaction is antiferromagnetic, whereas the interaction between the $t_{2g}^3$ and the empty $e_{g}^0$ is ferromagnetic.  In the monolayer CrI$_3$, the competing $t_{2g}^3-e_{g}^0$ superexchange dominates ensuring a ferromagnetic state for the system. Such multiorbital superexchange is complex, and the orbital occupancies play a critical role, which is affected by the charge doping. At a finite hole density $\delta_h$,  a partial hole $\delta_{\circ}$ is introduced at the $t_{2g}^{3}$ orbital, which destabilises the antiferromagnetic $t_{2g}-t_{2g}$ interactions. Therefore, a stronger ferromagnetism emerges with hole doping. Qualitatively, the $\delta_{\circ}$ further increases with hole doping, and the overall ferromagnetic exchange coupling gradually increases [Fig.~\ref{fig:figure4}(a)], which consequently leads to a gradual increase in the ferromagnetic ordering temperature (Fig.~\ref{fig:figure3}). The description for the electron doping follows a similar argument, and a partial charge $\delta_{\bullet}$ is introduced at the empty $e_{g}^0$ orbital at a finite $\delta_e$. This $\delta_{\bullet}$ destabilises the ferromagnetic $t_{2g}-e_{g}$ interaction, and the ferromagnetic ground state becomes weaker upon electron doping. Since $\delta_{\bullet}$ increases with electron doping, the resultant ferromagnetic exchange interaction decreases monotonically [Fig.~\ref{fig:figure4}(b)] and the ferromagnetic $T_{\C}$ gradually decreases (Fig.~\ref{fig:figure3}).

\begin{figure}[!t]
\begin{center}
{\includegraphics[width=0.49\textwidth, angle=0]{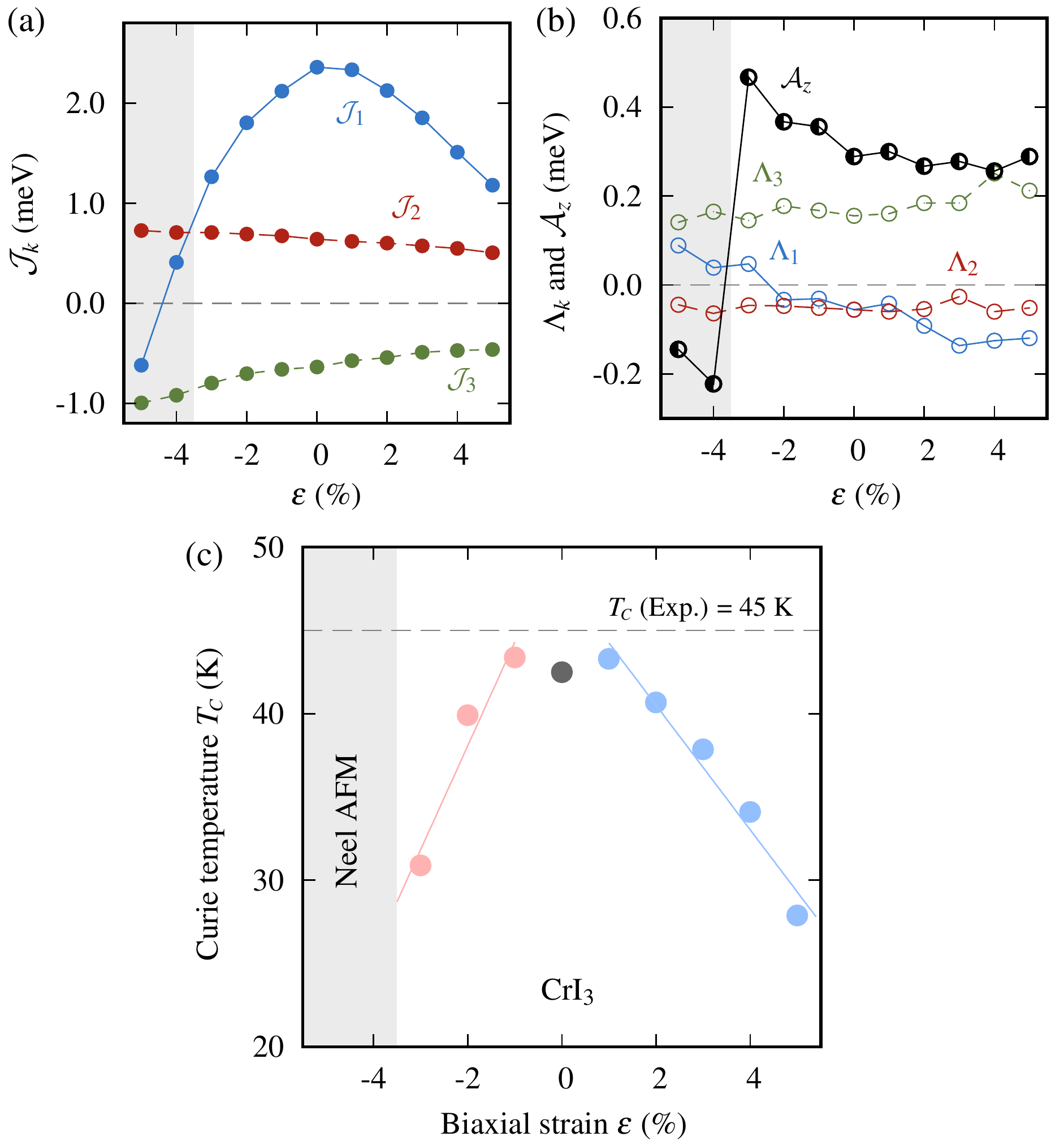}}
\caption{Biaxial strain $\varepsilon$ in the CrI$_3$ honeycomb lattice modifies the magnetic interactions and thus the ferromagnetic $T_{\C}$. (a)-(b) Variations of $\altmathcal{J}_k$, $\Lambda_k$ and $\altmathcal{A}_z$ on the compressive ($\varepsilon < 0)$ and tensile ($\varepsilon > 0)$ strains. While at a moderate strain the monolayer remains ferromagnetic, for a large compressive strain above $-$3\%, CrI$_3$ becomes antiferromagnetic with in-plane N\'{e}el order. (c) Calculated $T_{\C}$ gradually decreases under the influence of both compressive and tensile strains.  
}
\label{fig:figure5}
\end{center}
\end{figure}

It would be relevant to address the effects of strain while explaining the electrostatic doping dependent magnetism in CrI$_3$.  The lattice constants are modified under charge doping, and a biaxial tensile (compressive) strain is produced due to electron (hole) injection (Supplemental Material~\citep{supple}).  The induced strain ultimately perturbs the magnetic interactions in $\altmathcal{H}_{\raisebox{-1.5pt}{\footnotesize$\ell$}}$, and consequently affects the ordering temperature. Therefore, it would be appealing to decouple the contribution from the induced lattice strain itself. For the undoped CrI$_3$ monolayer, the ferromagnetic $\altmathcal{J}_1$ is strongly affected under the biaxial strain and monotonically decreases with both compressive $\varepsilon < 0$ and tensile $\varepsilon > 0$ strains [Fig.~\ref{fig:figure5}(a)]. While the ferromagnetic $\altmathcal{J}_2$ and antiferromagnetic $\altmathcal{J}_3$ exhibit a weaker dependence with increasing or decreasing lattice parameters. We also note that under a sizeable compressive strain beyond $-3$\%, the monolayer CrI$_3$ undergoes a magnetic phase transition and becomes antiferromagnetic with N\'{e}el order [Fig.~\ref{fig:figure5}(a)]. Further, the Cr-spins rearranges from out-of-plane to in-plane orientation with $\altmathcal{A}_z < 0$ [Fig.~\ref{fig:figure5}(b)].  Consolidating the microscopic variations of the magnetic interactions described in Fig.~\ref{fig:figure5}(a)-(b), we predict that $T_{\C}$ decreases under both compressive and tensile strains [Fig.~\ref{fig:figure5}(c)]. However, note that at a small strain range $-$1\% $< \varepsilon <$ 1\%, the $T_{\C}$ remains unaltered. Present results are consistent with the earlier efforts.~\citep{PhysRevB.98.144411}  It was also experimentally found that the hydrostatic pressure alters the interlayer magnetic coupling in thin CrI$_3$ flakes, and the $T_{\C}$ in bulk.~\citep{s41563-019-0505-2,s41563-019-0506-1} However, in contrast to the monolayer, the bulk CrI$_3$ behaves differently since the $T_{\C}$ was observed to increase with the hydrostatic pressure, and was explained to the modulation in interlayer exchange coupling.~\citep{PhysRevB.99.180407}  

Now we analyse the above results in the context of the strain induced by charge doping. A 2.7\% biaxial compressive strain is induced by 0.5 hole/CrI$_3$, and the corresponding $T_{\C}$ decreases to 33 K for the updoped monolayer with this value of strain. This observation is in contrast to the finding in Fig.~\ref{fig:figure3}, which indicates that the hole-induced increase in ordering temperature is entirely of electronic origin. A similar qualitative conclusion can be drawn for the tensile strain produced by electron doping. A maximum of 1\% tensile strain is induced with $\delta_e$ studied in Fig.~\ref{fig:figure3}, and at this strain level $T_{\C}$ remains unaffected [Fig.~\ref{fig:figure5}(c)].

Inspired by the present results on monolayer CrI$_3$ (Fig.~\ref{fig:figure3}), we investigate the effects of charge doping in CrGeTe$_3$. However, and to our surprise, we notice that the results are contrary to CrI$_3$.  The out-of-plane magnetism in the neutral CrGeTe$_3$ switches easily upon charge doping, and the Cr-moments prefer in-plane orientation ($\altmathcal{A}_z < 0$).  Unlike CrI$_3$, the critical density is order-of-magnitude lower, which is less than 0.025/CrGeTe$_3$ $\sim$ 1.27 $\times$ 10$^{13}$ cm$^{-2}$. The results indicate that the magnetism in monolayer CrGeTe$_3$ could be easily perturbed and becomes a 2D-XY system, resulting in suppressing ordering, which might explain the non-observation of magnetic ordering in the monolayer. We also conclude that a charge-induced increase in ordering temperature is not a generic feature in 2D ferromagnets, but chemical constituents are crucial.

Similar to charge doping, chemical doping may also tune the anisotropic magnetic couplings. We have previously discussed that since spin-orbit coupling strength at the anion-site drops from CrI$_3$ to CrCl$_3$; the magnetic anisotropies decline monotonically (I $\rightarrow$ Br $\rightarrow$ Cl), resulting in decreasing ordering temperature (Table~\ref{tab:table1}). This fact has also been experimentally confirmed in mixed halide compounds CrCl$_{3-x}$Br$_x$.~\citep{adma.201801325} Similarly, anisotropies can be modulated by doping at the cation-site. In this regard, we investigated Mo$_{0.25}$Cr$_{0.75}$I$_3$ and W$_{0.25}$Cr$_{0.75}$I$_3$ (Table~\ref{tab:table1}), where both Mo and W are isoelectronic to Cr and assumes identical oxidation and spin states. Indeed the single-ion anisotropy $\altmathcal{A}_z$ increases considerably in these materials (Table~\ref{tab:table1}). Compared to the pristine CrI$_3$ monolayer, the nature of the magnetic interactions are qualitatively same in these materials, but the magnitudes of $\altmathcal{J}_k$ and $\Lambda_k$ are altered due to doping-dependent lattice parameters. Compiling all the interactions in Table~\ref{tab:table1} into the Monte Carlo simulations reveal that while the $T_{\C}$ increases to 50.5 K in W$_{0.25}$Cr$_{0.75}$I$_3$, it decreases to 33 K for Mo$_{0.25}$Cr$_{0.75}$I$_3$. Therefore, we argue that the chemical doping is not a promising route toward achieving the elusive room-temperature ferromagnetism in 2D.

\section{Summary and outlook}
To describe magnetism in the 2D materials, we develop an anisotropic Heisenberg model with exchange interactions beyond the first-neighbour, which has been primarily neglected earlier.\citep{Lado_2017, Torelli_2018, s41524-018-0115-6, PhysRevB.98.144411, PhysRevB.100.205409, PhysRevB.101.134418, PhysRevMaterials.4.094004,PhysRevB.102.115162, PhysRevLett.124.017201} The parameters of the model, single-ion anisotropy, isotropic and anisotropic exchange interactions, are calculated from the relativistic first-principles calculations. We demonstrate that the second and third neighbouring magnetic interactions on a honeycomb lattice are critical to qualitatively and quantitatively explain the existing experimental results and trends.  The model incorporates various spin excitations on a honeycomb lattice and quantitatively reproduces the experimental $T_{\C}$ for the monolayer ferromagnetic insulators such as CrI$_3$,~\citep{nature22391} CrBr$_3$,~\citep{Kim11131, acs.nanolett.9b00553} and CrCl$_3$.~\citep{bedoyapinto2020intrinsic} For CrGeTe$_3$, while we predict a magnetic ordering in the monolayer, the experimental demonstration remains elusive. However, the layer-dependent $T_{\C}$~\citep{nature22060} and the present results indicate that a long-range ordering in the monolayer should exist, and therefore, a further experimental and theoretical investigations are necessary. Small on-site magnetic anisotropy in these materials triggers finite-temperature spin-fluctuations, and therefore the corresponding critical temperatures are much lower than the corresponding Ising picture. The magnetic critical exponent $\beta$ ranges between 0.21 to 0.29 for the monolayers, which are expectedly but only slightly lower than their bulk analogues.~\citep{PhysRevB.97.014420,PhysRevB.101.134418,s41928-019-0302-6,PhysRevB.102.014424,PhysRevB.95.245212,PhysRevB.98.214420}  This fact signifies a weaker interlayer magnetic coupling in the bulk and few-layer flakes; and the nature of magnetism grossly remains thickness-independent. Moreover, the calculated $\beta$ for the monolayers lie between the values corresponding to the 2D Ising ($\beta=$ 0.125) and 3D Heisenberg ($\beta=$ 0.365) models. Due to the presence of various magnetic anisotropies, the spin excitation becomes gapped, and the zero-temperature spin-wave gaps are in agreement with a range of measurements such as ferromagnetic resonance,~\citep{PhysRevLett.124.017201} inelastic neutron scattering experiment,~\citep{PhysRevB.101.134418} and magneto-Raman spectroscopy.~\citep{s41567-020-0999-1} While the present model has been adopted for the honeycomb lattice, it is readily applicable to the materials with different lattice symmetries.   

The predictive model described here can be further employed to investigate the impacts of various external perturbations and chemical doping. We illustrate that the charge doping critically affects the isotropic and anisotropic magnetic interactions that modifies the ordering temperature.  Moreover, hole doping considerably enhances the $T_{\C}$ and a five-fold increase to 228 K is witnessed at 2.66 $\times$ 10$^{-14}$ cm$^{-2}$ hole density.  Such a high carrier density is experimentally realizable in the 2D materials through electrostatic gating~\citep{ncomms9826,nnano.2015.314} or femtosecond laser pulse.~\citep{PhysRevLett.125.267205} In contrast, the electron doping has a detrimental effect on the ordering as $T_{\C}$ gradually decreases.  We notice that below a critical carrier density, the magnetism remains out-of-plane and beyond which it switches to the in-plane orientation as $\altmathcal{A}_z$ changes sign. While a similar qualitative trend is followed experimentally at a lower carrier density,~\citep{s41565-018-0135-x} here we also render the elusive microscopic mechanism for carrier tuneable magnetism (Fig.~\ref{fig:figure3}). We further describe the effects of induced biaxial strain in the honeycomb lattice and find that the compressive and tensile strains modify the magnetic interaction parameters such that in both cases, the ordering temperature decreases monotonically. Considering the trend of $T_{\C}$ in neutral CrI$_3$ under the biaxial strain, we conclude that carrier tuneable magnetism is entirely of electronic origin. Overall, while we argue that electrical manipulation of ordering temperature is a promising route to achieve room-temperature ferromagnetism in the 2D van der Walls materials, chemical doping seems not so encouraging. The present results should motivate further experimental studies in this regard.

\section{Acknowledgements}
M.K. acknowledges funding from the Indian Department of Science and Technology through Nano Mission project SR/NM/TP-13/2016, and the Science and Engineering Research Board through EMR/2016/006458 grant. We gratefully acknowledge the support and resources provided by the PARAM Brahma Facility at the Indian Institute of Science Education and Research, Pune under the National Supercomputing Mission of Government of India.


\begin{thebibliography}{78}%
\makeatletter
\providecommand \@ifxundefined [1]{%
 \@ifx{#1\undefined}
}%
\providecommand \@ifnum [1]{%
 \ifnum #1\expandafter \@firstoftwo
 \else \expandafter \@secondoftwo
 \fi
}%
\providecommand \@ifx [1]{%
 \ifx #1\expandafter \@firstoftwo
 \else \expandafter \@secondoftwo
 \fi
}%
\providecommand \natexlab [1]{#1}%
\providecommand \enquote  [1]{``#1''}%
\providecommand \bibnamefont  [1]{#1}%
\providecommand \bibfnamefont [1]{#1}%
\providecommand \citenamefont [1]{#1}%
\providecommand \href@noop [0]{\@secondoftwo}%
\providecommand \href [0]{\begingroup \@sanitize@url \@href}%
\providecommand \@href[1]{\@@startlink{#1}\@@href}%
\providecommand \@@href[1]{\endgroup#1\@@endlink}%
\providecommand \@sanitize@url [0]{\catcode `\\12\catcode `\$12\catcode
  `\&12\catcode `\#12\catcode `\^12\catcode `\_12\catcode `\%12\relax}%
\providecommand \@@startlink[1]{}%
\providecommand \@@endlink[0]{}%
\providecommand \url  [0]{\begingroup\@sanitize@url \@url }%
\providecommand \@url [1]{\endgroup\@href {#1}{\urlprefix }}%
\providecommand \urlprefix  [0]{URL }%
\providecommand \Eprint [0]{\href }%
\providecommand \doibase [0]{http://dx.doi.org/}%
\providecommand \selectlanguage [0]{\@gobble}%
\providecommand \bibinfo  [0]{\@secondoftwo}%
\providecommand \bibfield  [0]{\@secondoftwo}%
\providecommand \translation [1]{[#1]}%
\providecommand \BibitemOpen [0]{}%
\providecommand \bibitemStop [0]{}%
\providecommand \bibitemNoStop [0]{.\EOS\space}%
\providecommand \EOS [0]{\spacefactor3000\relax}%
\providecommand \BibitemShut  [1]{\csname bibitem#1\endcsname}%
\let\auto@bib@innerbib\@empty
\bibitem [{\citenamefont {Hohenberg}(1967)}]{PhysRev.158.383}%
  \BibitemOpen
  \bibfield  {author} {\bibinfo {author} {\bibfnamefont {P.~C.}\ \bibnamefont
  {Hohenberg}},\ }\href {https://link.aps.org/doi/10.1103/PhysRev.158.383}
  {\bibfield  {journal} {\bibinfo  {journal} {Phys. Rev.}\ }\textbf {\bibinfo
  {volume} {158}},\ \bibinfo {pages} {383} (\bibinfo {year}
  {1967})}\BibitemShut {NoStop}%
\bibitem [{\citenamefont {Mermin}\ and\ \citenamefont
  {Wagner}(1966)}]{PhysRevLett.17.1133}%
  \BibitemOpen
  \bibfield  {author} {\bibinfo {author} {\bibfnamefont {N.~D.}\ \bibnamefont
  {Mermin}}\ and\ \bibinfo {author} {\bibfnamefont {H.}~\bibnamefont
  {Wagner}},\ }\href {https://link.aps.org/doi/10.1103/PhysRevLett.17.1133}
  {\bibfield  {journal} {\bibinfo  {journal} {Phys. Rev. Lett.}\ }\textbf
  {\bibinfo {volume} {17}},\ \bibinfo {pages} {1133} (\bibinfo {year}
  {1966})}\BibitemShut {NoStop}%
\bibitem [{\citenamefont {Lee}\ \emph {et~al.}(2016)\citenamefont {Lee},
  \citenamefont {Lee}, \citenamefont {Ryoo}, \citenamefont {Kang},
  \citenamefont {Kim}, \citenamefont {Kim}, \citenamefont {Park}, \citenamefont
  {Park},\ and\ \citenamefont {Cheong}}]{nanolett.6b03052}%
  \BibitemOpen
  \bibfield  {author} {\bibinfo {author} {\bibfnamefont {J.-U.}\ \bibnamefont
  {Lee}}, \bibinfo {author} {\bibfnamefont {S.}~\bibnamefont {Lee}}, \bibinfo
  {author} {\bibfnamefont {J.~H.}\ \bibnamefont {Ryoo}}, \bibinfo {author}
  {\bibfnamefont {S.}~\bibnamefont {Kang}}, \bibinfo {author} {\bibfnamefont
  {T.~Y.}\ \bibnamefont {Kim}}, \bibinfo {author} {\bibfnamefont
  {P.}~\bibnamefont {Kim}}, \bibinfo {author} {\bibfnamefont {C.-H.}\
  \bibnamefont {Park}}, \bibinfo {author} {\bibfnamefont {J.-G.}\ \bibnamefont
  {Park}}, \ and\ \bibinfo {author} {\bibfnamefont {H.}~\bibnamefont
  {Cheong}},\ }\href {https://doi.org/10.1021/acs.nanolett.6b03052} {\bibfield
  {journal} {\bibinfo  {journal} {Nano Lett.}\ }\textbf {\bibinfo {volume}
  {16}},\ \bibinfo {pages} {7433} (\bibinfo {year} {2016})}\BibitemShut
  {NoStop}%
\bibitem [{\citenamefont {Gong}\ \emph {et~al.}(2017)\citenamefont {Gong},
  \citenamefont {Li}, \citenamefont {Li}, \citenamefont {Ji}, \citenamefont
  {Stern}, \citenamefont {Xia}, \citenamefont {Cao}, \citenamefont {Bao},
  \citenamefont {Wang}, \citenamefont {Wang}, \citenamefont {Qiu},
  \citenamefont {Cava}, \citenamefont {Louie}, \citenamefont {Xia},\ and\
  \citenamefont {Zhang}}]{nature22060}%
  \BibitemOpen
  \bibfield  {author} {\bibinfo {author} {\bibfnamefont {C.}~\bibnamefont
  {Gong}}, \bibinfo {author} {\bibfnamefont {L.}~\bibnamefont {Li}}, \bibinfo
  {author} {\bibfnamefont {Z.}~\bibnamefont {Li}}, \bibinfo {author}
  {\bibfnamefont {H.}~\bibnamefont {Ji}}, \bibinfo {author} {\bibfnamefont
  {A.}~\bibnamefont {Stern}}, \bibinfo {author} {\bibfnamefont
  {Y.}~\bibnamefont {Xia}}, \bibinfo {author} {\bibfnamefont {T.}~\bibnamefont
  {Cao}}, \bibinfo {author} {\bibfnamefont {W.}~\bibnamefont {Bao}}, \bibinfo
  {author} {\bibfnamefont {C.~e.}\ \bibnamefont {Wang}}, \bibinfo {author}
  {\bibfnamefont {Y.}~\bibnamefont {Wang}}, \bibinfo {author} {\bibfnamefont
  {Z.~Q.}\ \bibnamefont {Qiu}}, \bibinfo {author} {\bibfnamefont {R.~J.}\
  \bibnamefont {Cava}}, \bibinfo {author} {\bibfnamefont {S.~G.}\ \bibnamefont
  {Louie}}, \bibinfo {author} {\bibfnamefont {J.}~\bibnamefont {Xia}}, \ and\
  \bibinfo {author} {\bibfnamefont {X.}~\bibnamefont {Zhang}},\ }\href
  {https://doi.org/10.1038/nature22060} {\bibfield  {journal} {\bibinfo
  {journal} {Nature}\ }\textbf {\bibinfo {volume} {546}},\ \bibinfo {pages}
  {265} (\bibinfo {year} {2017})}\BibitemShut {NoStop}%
\bibitem [{\citenamefont {Huang}\ \emph {et~al.}(2017)\citenamefont {Huang},
  \citenamefont {Clark}, \citenamefont {Navarro-Moratalla}, \citenamefont
  {Klein}, \citenamefont {Cheng}, \citenamefont {Seyler}, \citenamefont
  {Zhong}, \citenamefont {Schmidgall}, \citenamefont {McGuire}, \citenamefont
  {Cobden}, \citenamefont {Yao}, \citenamefont {Xiao}, \citenamefont
  {Jarillo-Herrero},\ and\ \citenamefont {Xu}}]{nature22391}%
  \BibitemOpen
  \bibfield  {author} {\bibinfo {author} {\bibfnamefont {B.}~\bibnamefont
  {Huang}}, \bibinfo {author} {\bibfnamefont {G.}~\bibnamefont {Clark}},
  \bibinfo {author} {\bibfnamefont {E.}~\bibnamefont {Navarro-Moratalla}},
  \bibinfo {author} {\bibfnamefont {D.~R.}\ \bibnamefont {Klein}}, \bibinfo
  {author} {\bibfnamefont {R.}~\bibnamefont {Cheng}}, \bibinfo {author}
  {\bibfnamefont {K.~L.}\ \bibnamefont {Seyler}}, \bibinfo {author}
  {\bibfnamefont {D.}~\bibnamefont {Zhong}}, \bibinfo {author} {\bibfnamefont
  {E.}~\bibnamefont {Schmidgall}}, \bibinfo {author} {\bibfnamefont {M.~A.}\
  \bibnamefont {McGuire}}, \bibinfo {author} {\bibfnamefont {D.~H.}\
  \bibnamefont {Cobden}}, \bibinfo {author} {\bibfnamefont {W.}~\bibnamefont
  {Yao}}, \bibinfo {author} {\bibfnamefont {D.}~\bibnamefont {Xiao}}, \bibinfo
  {author} {\bibfnamefont {P.}~\bibnamefont {Jarillo-Herrero}}, \ and\ \bibinfo
  {author} {\bibfnamefont {X.}~\bibnamefont {Xu}},\ }\href
  {https://doi.org/10.1038/nature22391} {\bibfield  {journal} {\bibinfo
  {journal} {Nature}\ }\textbf {\bibinfo {volume} {546}},\ \bibinfo {pages}
  {270} (\bibinfo {year} {2017})}\BibitemShut {NoStop}%
\bibitem [{\citenamefont {Gibertini}\ \emph {et~al.}(2019)\citenamefont
  {Gibertini}, \citenamefont {Koperski}, \citenamefont {Morpurgo},\ and\
  \citenamefont {Novoselov}}]{Gibertini2019}%
  \BibitemOpen
  \bibfield  {author} {\bibinfo {author} {\bibfnamefont {M.}~\bibnamefont
  {Gibertini}}, \bibinfo {author} {\bibfnamefont {M.}~\bibnamefont {Koperski}},
  \bibinfo {author} {\bibfnamefont {A.~F.}\ \bibnamefont {Morpurgo}}, \ and\
  \bibinfo {author} {\bibfnamefont {K.~S.}\ \bibnamefont {Novoselov}},\ }\href
  {https://doi.org/10.1038/s41565-019-0438-6} {\bibfield  {journal} {\bibinfo
  {journal} {Nat. Nanotechnol.}\ }\textbf {\bibinfo {volume} {14}},\ \bibinfo
  {pages} {408} (\bibinfo {year} {2019})}\BibitemShut {NoStop}%
\bibitem [{\citenamefont {Onsager}(1944)}]{PhysRev.65.117}%
  \BibitemOpen
  \bibfield  {author} {\bibinfo {author} {\bibfnamefont {L.}~\bibnamefont
  {Onsager}},\ }\href {https://link.aps.org/doi/10.1103/PhysRev.65.117}
  {\bibfield  {journal} {\bibinfo  {journal} {Phys. Rev.}\ }\textbf {\bibinfo
  {volume} {65}},\ \bibinfo {pages} {117} (\bibinfo {year} {1944})}\BibitemShut
  {NoStop}%
\bibitem [{\citenamefont {Kubo}\ and\ \citenamefont
  {Kishi}(1988)}]{PhysRevLett.61.2585}%
  \BibitemOpen
  \bibfield  {author} {\bibinfo {author} {\bibfnamefont {K.}~\bibnamefont
  {Kubo}}\ and\ \bibinfo {author} {\bibfnamefont {T.}~\bibnamefont {Kishi}},\
  }\href {\doibase 10.1103/PhysRevLett.61.2585} {\bibfield  {journal} {\bibinfo
   {journal} {Phys. Rev. Lett.}\ }\textbf {\bibinfo {volume} {61}},\ \bibinfo
  {pages} {2585} (\bibinfo {year} {1988})}\BibitemShut {NoStop}%
\bibitem [{\citenamefont {Lado}\ and\ \citenamefont
  {Fern{\'{a}}ndez-Rossier}(2017)}]{Lado_2017}%
  \BibitemOpen
  \bibfield  {author} {\bibinfo {author} {\bibfnamefont {J.~L.}\ \bibnamefont
  {Lado}}\ and\ \bibinfo {author} {\bibfnamefont {J.}~\bibnamefont
  {Fern{\'{a}}ndez-Rossier}},\ }\href
  {https://doi.org/10.1088%2F2053-1583%2Faa75ed} {\bibfield  {journal}
  {\bibinfo  {journal} {2D Mater.}\ }\textbf {\bibinfo {volume} {4}},\ \bibinfo
  {pages} {035002} (\bibinfo {year} {2017})}\BibitemShut {NoStop}%
\bibitem [{\citenamefont {Torelli}\ and\ \citenamefont
  {Olsen}(2018)}]{Torelli_2018}%
  \BibitemOpen
  \bibfield  {author} {\bibinfo {author} {\bibfnamefont {D.}~\bibnamefont
  {Torelli}}\ and\ \bibinfo {author} {\bibfnamefont {T.}~\bibnamefont
  {Olsen}},\ }\href {https://doi.org/10.1088%2F2053-1583%2Faaf06d} {\bibfield
  {journal} {\bibinfo  {journal} {2D Mater.}\ }\textbf {\bibinfo {volume}
  {6}},\ \bibinfo {pages} {015028} (\bibinfo {year} {2018})}\BibitemShut
  {NoStop}%
\bibitem [{\citenamefont {Berezinskii}(1972)}]{Berezinskii1972}%
  \BibitemOpen
  \bibfield  {author} {\bibinfo {author} {\bibfnamefont {V.~.~L.}\ \bibnamefont
  {Berezinskii}},\ }\href
  {http://www.jetp.ac.ru/cgi-bin/e/index/e/34/3/p610?a=list} {\bibfield
  {journal} {\bibinfo  {journal} {Sov. Phys. JETP}\ }\textbf {\bibinfo {volume}
  {34}},\ \bibinfo {pages} {610} (\bibinfo {year} {1972})}\BibitemShut
  {NoStop}%
\bibitem [{\citenamefont {Kosterlitz}\ and\ \citenamefont
  {Thouless}(1973)}]{Kosterlitz_1973}%
  \BibitemOpen
  \bibfield  {author} {\bibinfo {author} {\bibfnamefont {J.~M.}\ \bibnamefont
  {Kosterlitz}}\ and\ \bibinfo {author} {\bibfnamefont {D.~J.}\ \bibnamefont
  {Thouless}},\ }\href {https://doi.org/10.1088%2F0022-3719%2F6%2F7%2F010}
  {\bibfield  {journal} {\bibinfo  {journal} {J. Phys. C}\ }\textbf {\bibinfo
  {volume} {6}},\ \bibinfo {pages} {1181} (\bibinfo {year} {1973})}\BibitemShut
  {NoStop}%
\bibitem [{\citenamefont {Poulopoulos}\ and\ \citenamefont
  {Baberschke}(1999)}]{Poulopoulos_1999}%
  \BibitemOpen
  \bibfield  {author} {\bibinfo {author} {\bibfnamefont {P.}~\bibnamefont
  {Poulopoulos}}\ and\ \bibinfo {author} {\bibfnamefont {K.}~\bibnamefont
  {Baberschke}},\ }\href {https://doi.org/10.1088%2F0953-8984%2F11%2F48%2F310}
  {\bibfield  {journal} {\bibinfo  {journal} {J. Phys.: Condens. Matter}\
  }\textbf {\bibinfo {volume} {11}},\ \bibinfo {pages} {9495} (\bibinfo {year}
  {1999})}\BibitemShut {NoStop}%
\bibitem [{\citenamefont {Vaz}\ \emph {et~al.}(2008)\citenamefont {Vaz},
  \citenamefont {Bland},\ and\ \citenamefont {Lauhoff}}]{Vaz_2008}%
  \BibitemOpen
  \bibfield  {author} {\bibinfo {author} {\bibfnamefont {C.~A.~F.}\
  \bibnamefont {Vaz}}, \bibinfo {author} {\bibfnamefont {J.~A.~C.}\
  \bibnamefont {Bland}}, \ and\ \bibinfo {author} {\bibfnamefont
  {G.}~\bibnamefont {Lauhoff}},\ }\href
  {https://doi.org/10.1088%2F0034-4885%2F71%2F5%2F056501} {\bibfield  {journal}
  {\bibinfo  {journal} {Rep. Prog. Phys.}\ }\textbf {\bibinfo {volume} {71}},\
  \bibinfo {pages} {056501} (\bibinfo {year} {2008})}\BibitemShut {NoStop}%
\bibitem [{\citenamefont {McGuire}(2017)}]{cryst7050121}%
  \BibitemOpen
  \bibfield  {author} {\bibinfo {author} {\bibfnamefont {M.~A.}\ \bibnamefont
  {McGuire}},\ }\href {https://doi.org/10.3390/cryst7050121} {\bibfield
  {journal} {\bibinfo  {journal} {Crystals}\ }\textbf {\bibinfo {volume} {7}},\
  \bibinfo {pages} {121} (\bibinfo {year} {2017})}\BibitemShut {NoStop}%
\bibitem [{\citenamefont {Bedoya-Pinto}\ \emph {et~al.}(2020)\citenamefont
  {Bedoya-Pinto}, \citenamefont {Ji}, \citenamefont {Pandeya}, \citenamefont
  {Gargiani}, \citenamefont {Valvidares}, \citenamefont {Sessi}, \citenamefont
  {Radu}, \citenamefont {Chang},\ and\ \citenamefont
  {Parkin}}]{bedoyapinto2020intrinsic}%
  \BibitemOpen
  \bibfield  {author} {\bibinfo {author} {\bibfnamefont {A.}~\bibnamefont
  {Bedoya-Pinto}}, \bibinfo {author} {\bibfnamefont {J.-R.}\ \bibnamefont
  {Ji}}, \bibinfo {author} {\bibfnamefont {A.}~\bibnamefont {Pandeya}},
  \bibinfo {author} {\bibfnamefont {P.}~\bibnamefont {Gargiani}}, \bibinfo
  {author} {\bibfnamefont {M.}~\bibnamefont {Valvidares}}, \bibinfo {author}
  {\bibfnamefont {P.}~\bibnamefont {Sessi}}, \bibinfo {author} {\bibfnamefont
  {F.}~\bibnamefont {Radu}}, \bibinfo {author} {\bibfnamefont {K.}~\bibnamefont
  {Chang}}, \ and\ \bibinfo {author} {\bibfnamefont {S.}~\bibnamefont
  {Parkin}},\ }\href@noop {} {\enquote {\bibinfo {title} {Intrinsic 2d-xy
  ferromagnetism in a van der waals monolayer},}\ } (\bibinfo {year} {2020}),\
  \Eprint {http://arxiv.org/abs/2006.07605} {arXiv:2006.07605
  [cond-mat.mes-hall]} \BibitemShut {NoStop}%
\bibitem [{\citenamefont {Kim}\ \emph {et~al.}(2019{\natexlab{a}})\citenamefont
  {Kim}, \citenamefont {Yang}, \citenamefont {Li}, \citenamefont {Jiang},
  \citenamefont {Jin}, \citenamefont {Tao}, \citenamefont {Nichols},
  \citenamefont {Sfigakis}, \citenamefont {Zhong}, \citenamefont {Li},
  \citenamefont {Tian}, \citenamefont {Cory}, \citenamefont {Miao},
  \citenamefont {Shan}, \citenamefont {Mak}, \citenamefont {Lei}, \citenamefont
  {Sun}, \citenamefont {Zhao},\ and\ \citenamefont {Tsen}}]{Kim11131}%
  \BibitemOpen
  \bibfield  {author} {\bibinfo {author} {\bibfnamefont {H.~H.}\ \bibnamefont
  {Kim}}, \bibinfo {author} {\bibfnamefont {B.}~\bibnamefont {Yang}}, \bibinfo
  {author} {\bibfnamefont {S.}~\bibnamefont {Li}}, \bibinfo {author}
  {\bibfnamefont {S.}~\bibnamefont {Jiang}}, \bibinfo {author} {\bibfnamefont
  {C.}~\bibnamefont {Jin}}, \bibinfo {author} {\bibfnamefont {Z.}~\bibnamefont
  {Tao}}, \bibinfo {author} {\bibfnamefont {G.}~\bibnamefont {Nichols}},
  \bibinfo {author} {\bibfnamefont {F.}~\bibnamefont {Sfigakis}}, \bibinfo
  {author} {\bibfnamefont {S.}~\bibnamefont {Zhong}}, \bibinfo {author}
  {\bibfnamefont {C.}~\bibnamefont {Li}}, \bibinfo {author} {\bibfnamefont
  {S.}~\bibnamefont {Tian}}, \bibinfo {author} {\bibfnamefont {D.~G.}\
  \bibnamefont {Cory}}, \bibinfo {author} {\bibfnamefont {G.-X.}\ \bibnamefont
  {Miao}}, \bibinfo {author} {\bibfnamefont {J.}~\bibnamefont {Shan}}, \bibinfo
  {author} {\bibfnamefont {K.~F.}\ \bibnamefont {Mak}}, \bibinfo {author}
  {\bibfnamefont {H.}~\bibnamefont {Lei}}, \bibinfo {author} {\bibfnamefont
  {K.}~\bibnamefont {Sun}}, \bibinfo {author} {\bibfnamefont {L.}~\bibnamefont
  {Zhao}}, \ and\ \bibinfo {author} {\bibfnamefont {A.~W.}\ \bibnamefont
  {Tsen}},\ }\href {https://www.pnas.org/content/116/23/11131} {\bibfield
  {journal} {\bibinfo  {journal} {Proc. Natl. Acad. Sci. USA}\ }\textbf
  {\bibinfo {volume} {116}},\ \bibinfo {pages} {11131} (\bibinfo {year}
  {2019}{\natexlab{a}})}\BibitemShut {NoStop}%
\bibitem [{\citenamefont {Zhang}\ \emph {et~al.}(2019)\citenamefont {Zhang},
  \citenamefont {Shang}, \citenamefont {Jiang}, \citenamefont {Rasmita},
  \citenamefont {Gao},\ and\ \citenamefont {Yu}}]{acs.nanolett.9b00553}%
  \BibitemOpen
  \bibfield  {author} {\bibinfo {author} {\bibfnamefont {Z.}~\bibnamefont
  {Zhang}}, \bibinfo {author} {\bibfnamefont {J.}~\bibnamefont {Shang}},
  \bibinfo {author} {\bibfnamefont {C.}~\bibnamefont {Jiang}}, \bibinfo
  {author} {\bibfnamefont {A.}~\bibnamefont {Rasmita}}, \bibinfo {author}
  {\bibfnamefont {W.}~\bibnamefont {Gao}}, \ and\ \bibinfo {author}
  {\bibfnamefont {T.}~\bibnamefont {Yu}},\ }\href
  {https://doi.org/10.1021/acs.nanolett.9b00553} {\bibfield  {journal}
  {\bibinfo  {journal} {Nano Lett.}\ }\textbf {\bibinfo {volume} {19}},\
  \bibinfo {pages} {3138} (\bibinfo {year} {2019})}\BibitemShut {NoStop}%
\bibitem [{\citenamefont {Wang}\ \emph {et~al.}(2019)\citenamefont {Wang},
  \citenamefont {Gibertini}, \citenamefont {Dumcenco}, \citenamefont
  {Taniguchi}, \citenamefont {Watanabe}, \citenamefont {Giannini},\ and\
  \citenamefont {Morpurgo}}]{s41565-019-0565-0}%
  \BibitemOpen
  \bibfield  {author} {\bibinfo {author} {\bibfnamefont {Z.}~\bibnamefont
  {Wang}}, \bibinfo {author} {\bibfnamefont {M.}~\bibnamefont {Gibertini}},
  \bibinfo {author} {\bibfnamefont {D.}~\bibnamefont {Dumcenco}}, \bibinfo
  {author} {\bibfnamefont {T.}~\bibnamefont {Taniguchi}}, \bibinfo {author}
  {\bibfnamefont {K.}~\bibnamefont {Watanabe}}, \bibinfo {author}
  {\bibfnamefont {E.}~\bibnamefont {Giannini}}, \ and\ \bibinfo {author}
  {\bibfnamefont {A.~F.}\ \bibnamefont {Morpurgo}},\ }\href
  {https://doi.org/10.1038/s41565-019-0565-0} {\bibfield  {journal} {\bibinfo
  {journal} {Nat. Nanotechnol.}\ }\textbf {\bibinfo {volume} {14}},\ \bibinfo
  {pages} {1116} (\bibinfo {year} {2019})}\BibitemShut {NoStop}%
\bibitem [{\citenamefont {Klein}\ \emph {et~al.}(2019)\citenamefont {Klein},
  \citenamefont {MacNeill}, \citenamefont {Song}, \citenamefont {Larson},
  \citenamefont {Fang}, \citenamefont {Xu}, \citenamefont {Ribeiro},
  \citenamefont {Canfield}, \citenamefont {Kaxiras}, \citenamefont {Comin},\
  and\ \citenamefont {Jarillo-Herrero}}]{s41567-019-0651-0}%
  \BibitemOpen
  \bibfield  {author} {\bibinfo {author} {\bibfnamefont {D.~R.}\ \bibnamefont
  {Klein}}, \bibinfo {author} {\bibfnamefont {D.}~\bibnamefont {MacNeill}},
  \bibinfo {author} {\bibfnamefont {Q.}~\bibnamefont {Song}}, \bibinfo {author}
  {\bibfnamefont {D.~T.}\ \bibnamefont {Larson}}, \bibinfo {author}
  {\bibfnamefont {S.}~\bibnamefont {Fang}}, \bibinfo {author} {\bibfnamefont
  {M.}~\bibnamefont {Xu}}, \bibinfo {author} {\bibfnamefont {R.~A.}\
  \bibnamefont {Ribeiro}}, \bibinfo {author} {\bibfnamefont {P.~C.}\
  \bibnamefont {Canfield}}, \bibinfo {author} {\bibfnamefont {E.}~\bibnamefont
  {Kaxiras}}, \bibinfo {author} {\bibfnamefont {R.}~\bibnamefont {Comin}}, \
  and\ \bibinfo {author} {\bibfnamefont {P.}~\bibnamefont {Jarillo-Herrero}},\
  }\href {https://doi.org/10.1038/s41567-019-0651-0} {\bibfield  {journal}
  {\bibinfo  {journal} {Nat. Phys.}\ }\textbf {\bibinfo {volume} {15}},\
  \bibinfo {pages} {1255} (\bibinfo {year} {2019})}\BibitemShut {NoStop}%
\bibitem [{\citenamefont {Kong}\ \emph {et~al.}(2019)\citenamefont {Kong},
  \citenamefont {Stolze}, \citenamefont {Timmons}, \citenamefont {Tao},
  \citenamefont {Ni}, \citenamefont {Guo}, \citenamefont {Yang}, \citenamefont
  {Prozorov},\ and\ \citenamefont {Cava}}]{adma.201808074}%
  \BibitemOpen
  \bibfield  {author} {\bibinfo {author} {\bibfnamefont {T.}~\bibnamefont
  {Kong}}, \bibinfo {author} {\bibfnamefont {K.}~\bibnamefont {Stolze}},
  \bibinfo {author} {\bibfnamefont {E.~I.}\ \bibnamefont {Timmons}}, \bibinfo
  {author} {\bibfnamefont {J.}~\bibnamefont {Tao}}, \bibinfo {author}
  {\bibfnamefont {D.}~\bibnamefont {Ni}}, \bibinfo {author} {\bibfnamefont
  {S.}~\bibnamefont {Guo}}, \bibinfo {author} {\bibfnamefont {Z.}~\bibnamefont
  {Yang}}, \bibinfo {author} {\bibfnamefont {R.}~\bibnamefont {Prozorov}}, \
  and\ \bibinfo {author} {\bibfnamefont {R.~J.}\ \bibnamefont {Cava}},\ }\href
  {https://onlinelibrary.wiley.com/doi/abs/10.1002/adma.201808074} {\bibfield
  {journal} {\bibinfo  {journal} {Adv. Mater.}\ }\textbf {\bibinfo {volume}
  {31}},\ \bibinfo {pages} {1808074} (\bibinfo {year} {2019})}\BibitemShut
  {NoStop}%
\bibitem [{\citenamefont {Bonilla}\ \emph {et~al.}(2018)\citenamefont
  {Bonilla}, \citenamefont {Kolekar}, \citenamefont {Ma}, \citenamefont {Diaz},
  \citenamefont {Kalappattil}, \citenamefont {Das}, \citenamefont {Eggers},
  \citenamefont {Gutierrez}, \citenamefont {Phan},\ and\ \citenamefont
  {Batzill}}]{s41565-018-0063-9}%
  \BibitemOpen
  \bibfield  {author} {\bibinfo {author} {\bibfnamefont {M.}~\bibnamefont
  {Bonilla}}, \bibinfo {author} {\bibfnamefont {S.}~\bibnamefont {Kolekar}},
  \bibinfo {author} {\bibfnamefont {Y.}~\bibnamefont {Ma}}, \bibinfo {author}
  {\bibfnamefont {H.~C.}\ \bibnamefont {Diaz}}, \bibinfo {author}
  {\bibfnamefont {V.}~\bibnamefont {Kalappattil}}, \bibinfo {author}
  {\bibfnamefont {R.}~\bibnamefont {Das}}, \bibinfo {author} {\bibfnamefont
  {T.}~\bibnamefont {Eggers}}, \bibinfo {author} {\bibfnamefont {H.~R.}\
  \bibnamefont {Gutierrez}}, \bibinfo {author} {\bibfnamefont {M.-H.}\
  \bibnamefont {Phan}}, \ and\ \bibinfo {author} {\bibfnamefont
  {M.}~\bibnamefont {Batzill}},\ }\href
  {https://doi.org/10.1038/s41565-018-0063-9} {\bibfield  {journal} {\bibinfo
  {journal} {Nat. Nanotechnol.}\ }\textbf {\bibinfo {volume} {13}},\ \bibinfo
  {pages} {289} (\bibinfo {year} {2018})}\BibitemShut {NoStop}%
\bibitem [{\citenamefont {Deng}\ \emph {et~al.}(2018)\citenamefont {Deng},
  \citenamefont {Yu}, \citenamefont {Song}, \citenamefont {Zhang},
  \citenamefont {Wang}, \citenamefont {Sun}, \citenamefont {Yi}, \citenamefont
  {Wu}, \citenamefont {Wu}, \citenamefont {Zhu}, \citenamefont {Wang},
  \citenamefont {Chen},\ and\ \citenamefont {Zhang}}]{s41586-018-0626-9}%
  \BibitemOpen
  \bibfield  {author} {\bibinfo {author} {\bibfnamefont {Y.}~\bibnamefont
  {Deng}}, \bibinfo {author} {\bibfnamefont {Y.}~\bibnamefont {Yu}}, \bibinfo
  {author} {\bibfnamefont {Y.}~\bibnamefont {Song}}, \bibinfo {author}
  {\bibfnamefont {J.}~\bibnamefont {Zhang}}, \bibinfo {author} {\bibfnamefont
  {N.~Z.}\ \bibnamefont {Wang}}, \bibinfo {author} {\bibfnamefont
  {Z.}~\bibnamefont {Sun}}, \bibinfo {author} {\bibfnamefont {Y.}~\bibnamefont
  {Yi}}, \bibinfo {author} {\bibfnamefont {Y.~Z.}\ \bibnamefont {Wu}}, \bibinfo
  {author} {\bibfnamefont {S.}~\bibnamefont {Wu}}, \bibinfo {author}
  {\bibfnamefont {J.}~\bibnamefont {Zhu}}, \bibinfo {author} {\bibfnamefont
  {J.}~\bibnamefont {Wang}}, \bibinfo {author} {\bibfnamefont {X.~H.}\
  \bibnamefont {Chen}}, \ and\ \bibinfo {author} {\bibfnamefont
  {Y.}~\bibnamefont {Zhang}},\ }\href
  {https://doi.org/10.1038/s41586-018-0626-9} {\bibfield  {journal} {\bibinfo
  {journal} {Nature}\ }\textbf {\bibinfo {volume} {563}},\ \bibinfo {pages}
  {94} (\bibinfo {year} {2018})}\BibitemShut {NoStop}%
\bibitem [{\citenamefont {Zhong}\ \emph {et~al.}(2017)\citenamefont {Zhong},
  \citenamefont {Seyler}, \citenamefont {Linpeng}, \citenamefont {Cheng},
  \citenamefont {Sivadas}, \citenamefont {Huang}, \citenamefont {Schmidgall},
  \citenamefont {Taniguchi}, \citenamefont {Watanabe}, \citenamefont {McGuire},
  \citenamefont {Yao}, \citenamefont {Xiao}, \citenamefont {Fu},\ and\
  \citenamefont {Xu}}]{Zhonge1603113}%
  \BibitemOpen
  \bibfield  {author} {\bibinfo {author} {\bibfnamefont {D.}~\bibnamefont
  {Zhong}}, \bibinfo {author} {\bibfnamefont {K.~L.}\ \bibnamefont {Seyler}},
  \bibinfo {author} {\bibfnamefont {X.}~\bibnamefont {Linpeng}}, \bibinfo
  {author} {\bibfnamefont {R.}~\bibnamefont {Cheng}}, \bibinfo {author}
  {\bibfnamefont {N.}~\bibnamefont {Sivadas}}, \bibinfo {author} {\bibfnamefont
  {B.}~\bibnamefont {Huang}}, \bibinfo {author} {\bibfnamefont
  {E.}~\bibnamefont {Schmidgall}}, \bibinfo {author} {\bibfnamefont
  {T.}~\bibnamefont {Taniguchi}}, \bibinfo {author} {\bibfnamefont
  {K.}~\bibnamefont {Watanabe}}, \bibinfo {author} {\bibfnamefont {M.~A.}\
  \bibnamefont {McGuire}}, \bibinfo {author} {\bibfnamefont {W.}~\bibnamefont
  {Yao}}, \bibinfo {author} {\bibfnamefont {D.}~\bibnamefont {Xiao}}, \bibinfo
  {author} {\bibfnamefont {K.-M.~C.}\ \bibnamefont {Fu}}, \ and\ \bibinfo
  {author} {\bibfnamefont {X.}~\bibnamefont {Xu}},\ }\href
  {https://advances.sciencemag.org/content/3/5/e1603113} {\bibfield  {journal}
  {\bibinfo  {journal} {Sci. Adv.}\ }\textbf {\bibinfo {volume} {3}} (\bibinfo
  {year} {2017})}\BibitemShut {NoStop}%
\bibitem [{\citenamefont {Song}\ \emph {et~al.}(2018)\citenamefont {Song},
  \citenamefont {Cai}, \citenamefont {Tu}, \citenamefont {Zhang}, \citenamefont
  {Huang}, \citenamefont {Wilson}, \citenamefont {Seyler}, \citenamefont {Zhu},
  \citenamefont {Taniguchi}, \citenamefont {Watanabe}, \citenamefont {McGuire},
  \citenamefont {Cobden}, \citenamefont {Xiao}, \citenamefont {Yao},\ and\
  \citenamefont {Xu}}]{Song1214}%
  \BibitemOpen
  \bibfield  {author} {\bibinfo {author} {\bibfnamefont {T.}~\bibnamefont
  {Song}}, \bibinfo {author} {\bibfnamefont {X.}~\bibnamefont {Cai}}, \bibinfo
  {author} {\bibfnamefont {M.~W.-Y.}\ \bibnamefont {Tu}}, \bibinfo {author}
  {\bibfnamefont {X.}~\bibnamefont {Zhang}}, \bibinfo {author} {\bibfnamefont
  {B.}~\bibnamefont {Huang}}, \bibinfo {author} {\bibfnamefont {N.~P.}\
  \bibnamefont {Wilson}}, \bibinfo {author} {\bibfnamefont {K.~L.}\
  \bibnamefont {Seyler}}, \bibinfo {author} {\bibfnamefont {L.}~\bibnamefont
  {Zhu}}, \bibinfo {author} {\bibfnamefont {T.}~\bibnamefont {Taniguchi}},
  \bibinfo {author} {\bibfnamefont {K.}~\bibnamefont {Watanabe}}, \bibinfo
  {author} {\bibfnamefont {M.~A.}\ \bibnamefont {McGuire}}, \bibinfo {author}
  {\bibfnamefont {D.~H.}\ \bibnamefont {Cobden}}, \bibinfo {author}
  {\bibfnamefont {D.}~\bibnamefont {Xiao}}, \bibinfo {author} {\bibfnamefont
  {W.}~\bibnamefont {Yao}}, \ and\ \bibinfo {author} {\bibfnamefont
  {X.}~\bibnamefont {Xu}},\ }\href
  {https://science.sciencemag.org/content/360/6394/1214} {\bibfield  {journal}
  {\bibinfo  {journal} {Science}\ }\textbf {\bibinfo {volume} {360}},\ \bibinfo
  {pages} {1214} (\bibinfo {year} {2018})}\BibitemShut {NoStop}%
\bibitem [{\citenamefont {Zhong}\ \emph {et~al.}(2020)\citenamefont {Zhong},
  \citenamefont {Seyler}, \citenamefont {Linpeng}, \citenamefont {Wilson},
  \citenamefont {Taniguchi}, \citenamefont {Watanabe}, \citenamefont {McGuire},
  \citenamefont {Fu}, \citenamefont {Xiao}, \citenamefont {Yao},\ and\
  \citenamefont {Xu}}]{s41565-019-0629-1}%
  \BibitemOpen
  \bibfield  {author} {\bibinfo {author} {\bibfnamefont {D.}~\bibnamefont
  {Zhong}}, \bibinfo {author} {\bibfnamefont {K.~L.}\ \bibnamefont {Seyler}},
  \bibinfo {author} {\bibfnamefont {X.}~\bibnamefont {Linpeng}}, \bibinfo
  {author} {\bibfnamefont {N.~P.}\ \bibnamefont {Wilson}}, \bibinfo {author}
  {\bibfnamefont {T.}~\bibnamefont {Taniguchi}}, \bibinfo {author}
  {\bibfnamefont {K.}~\bibnamefont {Watanabe}}, \bibinfo {author}
  {\bibfnamefont {M.~A.}\ \bibnamefont {McGuire}}, \bibinfo {author}
  {\bibfnamefont {K.-M.~C.}\ \bibnamefont {Fu}}, \bibinfo {author}
  {\bibfnamefont {D.}~\bibnamefont {Xiao}}, \bibinfo {author} {\bibfnamefont
  {W.}~\bibnamefont {Yao}}, \ and\ \bibinfo {author} {\bibfnamefont
  {X.}~\bibnamefont {Xu}},\ }\href {https://doi.org/10.1038/s41565-019-0629-1}
  {\bibfield  {journal} {\bibinfo  {journal} {Nat. Nanotechnol.}\ }\textbf
  {\bibinfo {volume} {15}},\ \bibinfo {pages} {187} (\bibinfo {year}
  {2020})}\BibitemShut {NoStop}%
\bibitem [{\citenamefont {Ciorciaro}\ \emph {et~al.}(2020)\citenamefont
  {Ciorciaro}, \citenamefont {Kroner}, \citenamefont {Watanabe}, \citenamefont
  {Taniguchi},\ and\ \citenamefont {Imamoglu}}]{PhysRevLett.124.197401}%
  \BibitemOpen
  \bibfield  {author} {\bibinfo {author} {\bibfnamefont {L.}~\bibnamefont
  {Ciorciaro}}, \bibinfo {author} {\bibfnamefont {M.}~\bibnamefont {Kroner}},
  \bibinfo {author} {\bibfnamefont {K.}~\bibnamefont {Watanabe}}, \bibinfo
  {author} {\bibfnamefont {T.}~\bibnamefont {Taniguchi}}, \ and\ \bibinfo
  {author} {\bibfnamefont {A.}~\bibnamefont {Imamoglu}},\ }\href
  {https://link.aps.org/doi/10.1103/PhysRevLett.124.197401} {\bibfield
  {journal} {\bibinfo  {journal} {Phys. Rev. Lett.}\ }\textbf {\bibinfo
  {volume} {124}},\ \bibinfo {pages} {197401} (\bibinfo {year}
  {2020})}\BibitemShut {NoStop}%
\bibitem [{\citenamefont {Kezilebieke}\ \emph {et~al.}(2020)\citenamefont
  {Kezilebieke}, \citenamefont {Huda}, \citenamefont {Va{\v n}o}, \citenamefont
  {Aapro}, \citenamefont {Ganguli}, \citenamefont {Silveira}, \citenamefont
  {G{\l}odzik}, \citenamefont {Foster}, \citenamefont {Ojanen},\ and\
  \citenamefont {Liljeroth}}]{kezilebieke2020topological}%
  \BibitemOpen
  \bibfield  {author} {\bibinfo {author} {\bibfnamefont {S.}~\bibnamefont
  {Kezilebieke}}, \bibinfo {author} {\bibfnamefont {M.~N.}\ \bibnamefont
  {Huda}}, \bibinfo {author} {\bibfnamefont {V.}~\bibnamefont {Va{\v n}o}},
  \bibinfo {author} {\bibfnamefont {M.}~\bibnamefont {Aapro}}, \bibinfo
  {author} {\bibfnamefont {S.~C.}\ \bibnamefont {Ganguli}}, \bibinfo {author}
  {\bibfnamefont {O.~J.}\ \bibnamefont {Silveira}}, \bibinfo {author}
  {\bibfnamefont {S.}~\bibnamefont {G{\l}odzik}}, \bibinfo {author}
  {\bibfnamefont {A.~S.}\ \bibnamefont {Foster}}, \bibinfo {author}
  {\bibfnamefont {T.}~\bibnamefont {Ojanen}}, \ and\ \bibinfo {author}
  {\bibfnamefont {P.}~\bibnamefont {Liljeroth}},\ }\href
  {https://doi.org/10.1038/s41586-020-2989-y} {\bibfield  {journal} {\bibinfo
  {journal} {Nature}\ }\textbf {\bibinfo {volume} {588}},\ \bibinfo {pages}
  {424} (\bibinfo {year} {2020})}\BibitemShut {NoStop}%
\bibitem [{\citenamefont {McGuire}\ \emph {et~al.}(2015)\citenamefont
  {McGuire}, \citenamefont {Dixit}, \citenamefont {Cooper},\ and\ \citenamefont
  {Sales}}]{cm504242t}%
  \BibitemOpen
  \bibfield  {author} {\bibinfo {author} {\bibfnamefont {M.~A.}\ \bibnamefont
  {McGuire}}, \bibinfo {author} {\bibfnamefont {H.}~\bibnamefont {Dixit}},
  \bibinfo {author} {\bibfnamefont {V.~R.}\ \bibnamefont {Cooper}}, \ and\
  \bibinfo {author} {\bibfnamefont {B.~C.}\ \bibnamefont {Sales}},\ }\href
  {https://doi.org/10.1021/cm504242t} {\bibfield  {journal} {\bibinfo
  {journal} {Chem. Mater.}\ }\textbf {\bibinfo {volume} {27}},\ \bibinfo
  {pages} {612} (\bibinfo {year} {2015})}\BibitemShut {NoStop}%
\bibitem [{\citenamefont {Carteaux}\ \emph {et~al.}(1995)\citenamefont
  {Carteaux}, \citenamefont {Brunet}, \citenamefont {Ouvrard},\ and\
  \citenamefont {Andre}}]{Carteaux_1995}%
  \BibitemOpen
  \bibfield  {author} {\bibinfo {author} {\bibfnamefont {V.}~\bibnamefont
  {Carteaux}}, \bibinfo {author} {\bibfnamefont {D.}~\bibnamefont {Brunet}},
  \bibinfo {author} {\bibfnamefont {G.}~\bibnamefont {Ouvrard}}, \ and\
  \bibinfo {author} {\bibfnamefont {G.}~\bibnamefont {Andre}},\ }\href
  {https://doi.org/10.1088%2F0953-8984%2F7%2F1%2F008} {\bibfield  {journal}
  {\bibinfo  {journal} {J. Phys.: Conden. Matter}\ }\textbf {\bibinfo {volume}
  {7}},\ \bibinfo {pages} {69} (\bibinfo {year} {1995})}\BibitemShut {NoStop}%
\bibitem [{\citenamefont {Anderson}(1950)}]{PhysRev.79.350}%
  \BibitemOpen
  \bibfield  {author} {\bibinfo {author} {\bibfnamefont {P.~W.}\ \bibnamefont
  {Anderson}},\ }\href {https://link.aps.org/doi/10.1103/PhysRev.79.350}
  {\bibfield  {journal} {\bibinfo  {journal} {Phys. Rev.}\ }\textbf {\bibinfo
  {volume} {79}},\ \bibinfo {pages} {350} (\bibinfo {year} {1950})}\BibitemShut
  {NoStop}%
\bibitem [{\citenamefont {Goodenough}(1958)}]{GOODENOUGH1958287}%
  \BibitemOpen
  \bibfield  {author} {\bibinfo {author} {\bibfnamefont {J.~B.}\ \bibnamefont
  {Goodenough}},\ }\href
  {http://www.sciencedirect.com/science/article/pii/0022369758901070}
  {\bibfield  {journal} {\bibinfo  {journal} {J. Phys. Chem. Solids}\ }\textbf
  {\bibinfo {volume} {6}},\ \bibinfo {pages} {287 } (\bibinfo {year}
  {1958})}\BibitemShut {NoStop}%
\bibitem [{\citenamefont {Kanamori}(1959)}]{KANAMORI195987}%
  \BibitemOpen
  \bibfield  {author} {\bibinfo {author} {\bibfnamefont {J.}~\bibnamefont
  {Kanamori}},\ }\href
  {http://www.sciencedirect.com/science/article/pii/0022369759900617}
  {\bibfield  {journal} {\bibinfo  {journal} {J. Phys. Chem. Solids}\ }\textbf
  {\bibinfo {volume} {10}},\ \bibinfo {pages} {87 } (\bibinfo {year}
  {1959})}\BibitemShut {NoStop}%
\bibitem [{\citenamefont {Webster}\ and\ \citenamefont
  {Yan}(2018)}]{PhysRevB.98.144411}%
  \BibitemOpen
  \bibfield  {author} {\bibinfo {author} {\bibfnamefont {L.}~\bibnamefont
  {Webster}}\ and\ \bibinfo {author} {\bibfnamefont {J.-A.}\ \bibnamefont
  {Yan}},\ }\href {https://link.aps.org/doi/10.1103/PhysRevB.98.144411}
  {\bibfield  {journal} {\bibinfo  {journal} {Phys. Rev. B}\ }\textbf {\bibinfo
  {volume} {98}},\ \bibinfo {pages} {144411} (\bibinfo {year}
  {2018})}\BibitemShut {NoStop}%
\bibitem [{\citenamefont {Xu}\ \emph {et~al.}(2018)\citenamefont {Xu},
  \citenamefont {Feng}, \citenamefont {Xiang},\ and\ \citenamefont
  {Bellaiche}}]{s41524-018-0115-6}%
  \BibitemOpen
  \bibfield  {author} {\bibinfo {author} {\bibfnamefont {C.}~\bibnamefont
  {Xu}}, \bibinfo {author} {\bibfnamefont {J.}~\bibnamefont {Feng}}, \bibinfo
  {author} {\bibfnamefont {H.}~\bibnamefont {Xiang}}, \ and\ \bibinfo {author}
  {\bibfnamefont {L.}~\bibnamefont {Bellaiche}},\ }\href
  {https://doi.org/10.1038/s41524-018-0115-6} {\bibfield  {journal} {\bibinfo
  {journal} {npj Comput. Mater.}\ }\textbf {\bibinfo {volume} {4}},\ \bibinfo
  {pages} {57} (\bibinfo {year} {2018})}\BibitemShut {NoStop}%
\bibitem [{\citenamefont {Lee}\ \emph {et~al.}(2020)\citenamefont {Lee},
  \citenamefont {Utermohlen}, \citenamefont {Weber}, \citenamefont {Hwang},
  \citenamefont {Zhang}, \citenamefont {van Tol}, \citenamefont {Goldberger},
  \citenamefont {Trivedi},\ and\ \citenamefont
  {Hammel}}]{PhysRevLett.124.017201}%
  \BibitemOpen
  \bibfield  {author} {\bibinfo {author} {\bibfnamefont {I.}~\bibnamefont
  {Lee}}, \bibinfo {author} {\bibfnamefont {F.~G.}\ \bibnamefont {Utermohlen}},
  \bibinfo {author} {\bibfnamefont {D.}~\bibnamefont {Weber}}, \bibinfo
  {author} {\bibfnamefont {K.}~\bibnamefont {Hwang}}, \bibinfo {author}
  {\bibfnamefont {C.}~\bibnamefont {Zhang}}, \bibinfo {author} {\bibfnamefont
  {J.}~\bibnamefont {van Tol}}, \bibinfo {author} {\bibfnamefont {J.~E.}\
  \bibnamefont {Goldberger}}, \bibinfo {author} {\bibfnamefont
  {N.}~\bibnamefont {Trivedi}}, \ and\ \bibinfo {author} {\bibfnamefont
  {P.~C.}\ \bibnamefont {Hammel}},\ }\href
  {https://link.aps.org/doi/10.1103/PhysRevLett.124.017201} {\bibfield
  {journal} {\bibinfo  {journal} {Phys. Rev. Lett.}\ }\textbf {\bibinfo
  {volume} {124}},\ \bibinfo {pages} {017201} (\bibinfo {year}
  {2020})}\BibitemShut {NoStop}%
\bibitem [{\citenamefont {Chen}\ \emph {et~al.}(2020)\citenamefont {Chen},
  \citenamefont {Chung}, \citenamefont {Chen}, \citenamefont {Duan},
  \citenamefont {Schneidewind}, \citenamefont {Radelytskyi}, \citenamefont
  {Voneshen}, \citenamefont {Ewings}, \citenamefont {Stone}, \citenamefont
  {Kolesnikov}, \citenamefont {Winn}, \citenamefont {Chi}, \citenamefont
  {Mole}, \citenamefont {Yu}, \citenamefont {Gao},\ and\ \citenamefont
  {Dai}}]{PhysRevB.101.134418}%
  \BibitemOpen
  \bibfield  {author} {\bibinfo {author} {\bibfnamefont {L.}~\bibnamefont
  {Chen}}, \bibinfo {author} {\bibfnamefont {J.-H.}\ \bibnamefont {Chung}},
  \bibinfo {author} {\bibfnamefont {T.}~\bibnamefont {Chen}}, \bibinfo {author}
  {\bibfnamefont {C.}~\bibnamefont {Duan}}, \bibinfo {author} {\bibfnamefont
  {A.}~\bibnamefont {Schneidewind}}, \bibinfo {author} {\bibfnamefont
  {I.}~\bibnamefont {Radelytskyi}}, \bibinfo {author} {\bibfnamefont {D.~J.}\
  \bibnamefont {Voneshen}}, \bibinfo {author} {\bibfnamefont {R.~A.}\
  \bibnamefont {Ewings}}, \bibinfo {author} {\bibfnamefont {M.~B.}\
  \bibnamefont {Stone}}, \bibinfo {author} {\bibfnamefont {A.~I.}\ \bibnamefont
  {Kolesnikov}}, \bibinfo {author} {\bibfnamefont {B.}~\bibnamefont {Winn}},
  \bibinfo {author} {\bibfnamefont {S.}~\bibnamefont {Chi}}, \bibinfo {author}
  {\bibfnamefont {R.~A.}\ \bibnamefont {Mole}}, \bibinfo {author}
  {\bibfnamefont {D.~H.}\ \bibnamefont {Yu}}, \bibinfo {author} {\bibfnamefont
  {B.}~\bibnamefont {Gao}}, \ and\ \bibinfo {author} {\bibfnamefont
  {P.}~\bibnamefont {Dai}},\ }\href
  {https://link.aps.org/doi/10.1103/PhysRevB.101.134418} {\bibfield  {journal}
  {\bibinfo  {journal} {Phys. Rev. B}\ }\textbf {\bibinfo {volume} {101}},\
  \bibinfo {pages} {134418} (\bibinfo {year} {2020})}\BibitemShut {NoStop}%
\bibitem [{\citenamefont {Vishkayi}\ \emph {et~al.}(2020)\citenamefont
  {Vishkayi}, \citenamefont {Torbatian}, \citenamefont {Qaiumzadeh},\ and\
  \citenamefont {Asgari}}]{PhysRevMaterials.4.094004}%
  \BibitemOpen
  \bibfield  {author} {\bibinfo {author} {\bibfnamefont {S.~I.}\ \bibnamefont
  {Vishkayi}}, \bibinfo {author} {\bibfnamefont {Z.}~\bibnamefont {Torbatian}},
  \bibinfo {author} {\bibfnamefont {A.}~\bibnamefont {Qaiumzadeh}}, \ and\
  \bibinfo {author} {\bibfnamefont {R.}~\bibnamefont {Asgari}},\ }\href
  {https://link.aps.org/doi/10.1103/PhysRevMaterials.4.094004} {\bibfield
  {journal} {\bibinfo  {journal} {Phys. Rev. Materials}\ }\textbf {\bibinfo
  {volume} {4}},\ \bibinfo {pages} {094004} (\bibinfo {year}
  {2020})}\BibitemShut {NoStop}%
\bibitem [{\citenamefont {Lu}\ \emph {et~al.}(2019)\citenamefont {Lu},
  \citenamefont {Fei},\ and\ \citenamefont {Yang}}]{PhysRevB.100.205409}%
  \BibitemOpen
  \bibfield  {author} {\bibinfo {author} {\bibfnamefont {X.}~\bibnamefont
  {Lu}}, \bibinfo {author} {\bibfnamefont {R.}~\bibnamefont {Fei}}, \ and\
  \bibinfo {author} {\bibfnamefont {L.}~\bibnamefont {Yang}},\ }\href
  {https://link.aps.org/doi/10.1103/PhysRevB.100.205409} {\bibfield  {journal}
  {\bibinfo  {journal} {Phys. Rev. B}\ }\textbf {\bibinfo {volume} {100}},\
  \bibinfo {pages} {205409} (\bibinfo {year} {2019})}\BibitemShut {NoStop}%
\bibitem [{\citenamefont {Kvashnin}\ \emph {et~al.}(2020)\citenamefont
  {Kvashnin}, \citenamefont {Bergman}, \citenamefont {Lichtenstein},\ and\
  \citenamefont {Katsnelson}}]{PhysRevB.102.115162}%
  \BibitemOpen
  \bibfield  {author} {\bibinfo {author} {\bibfnamefont {Y.~O.}\ \bibnamefont
  {Kvashnin}}, \bibinfo {author} {\bibfnamefont {A.}~\bibnamefont {Bergman}},
  \bibinfo {author} {\bibfnamefont {A.~I.}\ \bibnamefont {Lichtenstein}}, \
  and\ \bibinfo {author} {\bibfnamefont {M.~I.}\ \bibnamefont {Katsnelson}},\
  }\href {https://link.aps.org/doi/10.1103/PhysRevB.102.115162} {\bibfield
  {journal} {\bibinfo  {journal} {Phys. Rev. B}\ }\textbf {\bibinfo {volume}
  {102}},\ \bibinfo {pages} {115162} (\bibinfo {year} {2020})}\BibitemShut
  {NoStop}%
\bibitem [{\citenamefont {Owerre}(2016)}]{Owerre_2016}%
  \BibitemOpen
  \bibfield  {author} {\bibinfo {author} {\bibfnamefont {S.~A.}\ \bibnamefont
  {Owerre}},\ }\href {https://doi.org/10.1088/0953-8984/28/38/386001}
  {\bibfield  {journal} {\bibinfo  {journal} {J. Phys.: Condens. Matter}\
  }\textbf {\bibinfo {volume} {28}},\ \bibinfo {pages} {386001} (\bibinfo
  {year} {2016})}\BibitemShut {NoStop}%
\bibitem [{\citenamefont {Cheng}\ \emph {et~al.}(2016)\citenamefont {Cheng},
  \citenamefont {Okamoto},\ and\ \citenamefont
  {Xiao}}]{PhysRevLett.117.217202}%
  \BibitemOpen
  \bibfield  {author} {\bibinfo {author} {\bibfnamefont {R.}~\bibnamefont
  {Cheng}}, \bibinfo {author} {\bibfnamefont {S.}~\bibnamefont {Okamoto}}, \
  and\ \bibinfo {author} {\bibfnamefont {D.}~\bibnamefont {Xiao}},\ }\href
  {https://link.aps.org/doi/10.1103/PhysRevLett.117.217202} {\bibfield
  {journal} {\bibinfo  {journal} {Phys. Rev. Lett.}\ }\textbf {\bibinfo
  {volume} {117}},\ \bibinfo {pages} {217202} (\bibinfo {year}
  {2016})}\BibitemShut {NoStop}%
\bibitem [{\citenamefont {Kim}\ \emph {et~al.}(2016)\citenamefont {Kim},
  \citenamefont {Ochoa}, \citenamefont {Zarzuela},\ and\ \citenamefont
  {Tserkovnyak}}]{PhysRevLett.117.227201}%
  \BibitemOpen
  \bibfield  {author} {\bibinfo {author} {\bibfnamefont {S.~K.}\ \bibnamefont
  {Kim}}, \bibinfo {author} {\bibfnamefont {H.}~\bibnamefont {Ochoa}}, \bibinfo
  {author} {\bibfnamefont {R.}~\bibnamefont {Zarzuela}}, \ and\ \bibinfo
  {author} {\bibfnamefont {Y.}~\bibnamefont {Tserkovnyak}},\ }\href
  {https://link.aps.org/doi/10.1103/PhysRevLett.117.227201} {\bibfield
  {journal} {\bibinfo  {journal} {Phys. Rev. Lett.}\ }\textbf {\bibinfo
  {volume} {117}},\ \bibinfo {pages} {227201} (\bibinfo {year}
  {2016})}\BibitemShut {NoStop}%
\bibitem [{\citenamefont {Chen}\ \emph {et~al.}(2018)\citenamefont {Chen},
  \citenamefont {Chung}, \citenamefont {Gao}, \citenamefont {Chen},
  \citenamefont {Stone}, \citenamefont {Kolesnikov}, \citenamefont {Huang},\
  and\ \citenamefont {Dai}}]{PhysRevX.8.041028}%
  \BibitemOpen
  \bibfield  {author} {\bibinfo {author} {\bibfnamefont {L.}~\bibnamefont
  {Chen}}, \bibinfo {author} {\bibfnamefont {J.-H.}\ \bibnamefont {Chung}},
  \bibinfo {author} {\bibfnamefont {B.}~\bibnamefont {Gao}}, \bibinfo {author}
  {\bibfnamefont {T.}~\bibnamefont {Chen}}, \bibinfo {author} {\bibfnamefont
  {M.~B.}\ \bibnamefont {Stone}}, \bibinfo {author} {\bibfnamefont {A.~I.}\
  \bibnamefont {Kolesnikov}}, \bibinfo {author} {\bibfnamefont
  {Q.}~\bibnamefont {Huang}}, \ and\ \bibinfo {author} {\bibfnamefont
  {P.}~\bibnamefont {Dai}},\ }\href
  {https://link.aps.org/doi/10.1103/PhysRevX.8.041028} {\bibfield  {journal}
  {\bibinfo  {journal} {Phys. Rev. X}\ }\textbf {\bibinfo {volume} {8}},\
  \bibinfo {pages} {041028} (\bibinfo {year} {2018})}\BibitemShut {NoStop}%
\bibitem [{\citenamefont {Wang}\ \emph {et~al.}(2016)\citenamefont {Wang},
  \citenamefont {Fan}, \citenamefont {Zhu},\ and\ \citenamefont
  {Wu}}]{Wang_2016}%
  \BibitemOpen
  \bibfield  {author} {\bibinfo {author} {\bibfnamefont {H.}~\bibnamefont
  {Wang}}, \bibinfo {author} {\bibfnamefont {F.}~\bibnamefont {Fan}}, \bibinfo
  {author} {\bibfnamefont {S.}~\bibnamefont {Zhu}}, \ and\ \bibinfo {author}
  {\bibfnamefont {H.}~\bibnamefont {Wu}},\ }\href
  {https://doi.org/10.1209/0295-5075/114/47001} {\bibfield  {journal} {\bibinfo
   {journal} {Europhys. Lett.}\ }\textbf {\bibinfo {volume} {114}},\ \bibinfo
  {pages} {47001} (\bibinfo {year} {2016})}\BibitemShut {NoStop}%
\bibitem [{\citenamefont {Kresse}\ and\ \citenamefont
  {Hafner}(1993)}]{PhysRevB.47.558}%
  \BibitemOpen
  \bibfield  {author} {\bibinfo {author} {\bibfnamefont {G.}~\bibnamefont
  {Kresse}}\ and\ \bibinfo {author} {\bibfnamefont {J.}~\bibnamefont
  {Hafner}},\ }\href {http://link.aps.org/doi/10.1103/PhysRevB.47.558}
  {\bibfield  {journal} {\bibinfo  {journal} {Phys. Rev. B}\ }\textbf {\bibinfo
  {volume} {47}},\ \bibinfo {pages} {558} (\bibinfo {year} {1993})}\BibitemShut
  {NoStop}%
\bibitem [{\citenamefont {Kresse}\ and\ \citenamefont
  {Furthm\"uller}(1996)}]{PhysRevB.54.11169}%
  \BibitemOpen
  \bibfield  {author} {\bibinfo {author} {\bibfnamefont {G.}~\bibnamefont
  {Kresse}}\ and\ \bibinfo {author} {\bibfnamefont {J.}~\bibnamefont
  {Furthm\"uller}},\ }\href {http://link.aps.org/doi/10.1103/PhysRevB.54.11169}
  {\bibfield  {journal} {\bibinfo  {journal} {Phys. Rev. B}\ }\textbf {\bibinfo
  {volume} {54}},\ \bibinfo {pages} {11169} (\bibinfo {year}
  {1996})}\BibitemShut {NoStop}%
\bibitem [{\citenamefont {Bl\"ochl}(1994)}]{PhysRevB.50.17953}%
  \BibitemOpen
  \bibfield  {author} {\bibinfo {author} {\bibfnamefont {P.~E.}\ \bibnamefont
  {Bl\"ochl}},\ }\href {http://link.aps.org/doi/10.1103/PhysRevB.50.17953}
  {\bibfield  {journal} {\bibinfo  {journal} {Phys. Rev. B}\ }\textbf {\bibinfo
  {volume} {50}},\ \bibinfo {pages} {17953} (\bibinfo {year}
  {1994})}\BibitemShut {NoStop}%
\bibitem [{\citenamefont {Dudarev}\ \emph {et~al.}(1998)\citenamefont
  {Dudarev}, \citenamefont {Botton}, \citenamefont {Savrasov}, \citenamefont
  {Humphreys},\ and\ \citenamefont {Sutton}}]{PhysRevB.57.1505}%
  \BibitemOpen
  \bibfield  {author} {\bibinfo {author} {\bibfnamefont {S.~L.}\ \bibnamefont
  {Dudarev}}, \bibinfo {author} {\bibfnamefont {G.~A.}\ \bibnamefont {Botton}},
  \bibinfo {author} {\bibfnamefont {S.~Y.}\ \bibnamefont {Savrasov}}, \bibinfo
  {author} {\bibfnamefont {C.~J.}\ \bibnamefont {Humphreys}}, \ and\ \bibinfo
  {author} {\bibfnamefont {A.~P.}\ \bibnamefont {Sutton}},\ }\href
  {http://link.aps.org/doi/10.1103/PhysRevB.57.1505} {\bibfield  {journal}
  {\bibinfo  {journal} {Phys. Rev. B}\ }\textbf {\bibinfo {volume} {57}},\
  \bibinfo {pages} {1505} (\bibinfo {year} {1998})}\BibitemShut {NoStop}%
\bibitem [{\citenamefont {Monkhorst}\ and\ \citenamefont
  {Pack}(1976)}]{PhysRevB.13.5188}%
  \BibitemOpen
  \bibfield  {author} {\bibinfo {author} {\bibfnamefont {H.~J.}\ \bibnamefont
  {Monkhorst}}\ and\ \bibinfo {author} {\bibfnamefont {J.~D.}\ \bibnamefont
  {Pack}},\ }\href {https://link.aps.org/doi/10.1103/PhysRevB.13.5188}
  {\bibfield  {journal} {\bibinfo  {journal} {Phys. Rev. B}\ }\textbf {\bibinfo
  {volume} {13}},\ \bibinfo {pages} {5188} (\bibinfo {year}
  {1976})}\BibitemShut {NoStop}%
\bibitem [{\citenamefont {Metropolis}\ \emph {et~al.}(1953)\citenamefont
  {Metropolis}, \citenamefont {Rosenbluth}, \citenamefont {Rosenbluth},
  \citenamefont {Teller},\ and\ \citenamefont {Teller}}]{10.1063/1.1699114}%
  \BibitemOpen
  \bibfield  {author} {\bibinfo {author} {\bibfnamefont {N.}~\bibnamefont
  {Metropolis}}, \bibinfo {author} {\bibfnamefont {A.~W.}\ \bibnamefont
  {Rosenbluth}}, \bibinfo {author} {\bibfnamefont {M.~N.}\ \bibnamefont
  {Rosenbluth}}, \bibinfo {author} {\bibfnamefont {A.~H.}\ \bibnamefont
  {Teller}}, \ and\ \bibinfo {author} {\bibfnamefont {E.}~\bibnamefont
  {Teller}},\ }\href {https://doi.org/10.1063/1.1699114} {\bibfield  {journal}
  {\bibinfo  {journal} {J. Chem. Phys.}\ }\textbf {\bibinfo {volume} {21}},\
  \bibinfo {pages} {1087} (\bibinfo {year} {1953})}\BibitemShut {NoStop}%
\bibitem [{\citenamefont {Marsaglia}(1972)}]{marsaglia1972}%
  \BibitemOpen
  \bibfield  {author} {\bibinfo {author} {\bibfnamefont {G.}~\bibnamefont
  {Marsaglia}},\ }\href {https://doi.org/10.1214/aoms/1177692644} {\bibfield
  {journal} {\bibinfo  {journal} {Ann. Math. Statist.}\ }\textbf {\bibinfo
  {volume} {43}},\ \bibinfo {pages} {645} (\bibinfo {year} {1972})}\BibitemShut
  {NoStop}%
\bibitem [{sup()}]{supple}%
  \BibitemOpen
  \href@noop {} {}\bibinfo {note} {Supplemental Material.}\BibitemShut {Stop}%
\bibitem [{\citenamefont {Wildes}\ \emph {et~al.}(2015)\citenamefont {Wildes},
  \citenamefont {Simonet}, \citenamefont {Ressouche}, \citenamefont {McIntyre},
  \citenamefont {Avdeev}, \citenamefont {Suard}, \citenamefont {Kimber},
  \citenamefont {Lan\ifmmode~\mbox{\c{c}}\else \c{c}\fi{}on}, \citenamefont
  {Pepe}, \citenamefont {Moubaraki},\ and\ \citenamefont
  {Hicks}}]{PhysRevB.92.224408}%
  \BibitemOpen
  \bibfield  {author} {\bibinfo {author} {\bibfnamefont {A.~R.}\ \bibnamefont
  {Wildes}}, \bibinfo {author} {\bibfnamefont {V.}~\bibnamefont {Simonet}},
  \bibinfo {author} {\bibfnamefont {E.}~\bibnamefont {Ressouche}}, \bibinfo
  {author} {\bibfnamefont {G.~J.}\ \bibnamefont {McIntyre}}, \bibinfo {author}
  {\bibfnamefont {M.}~\bibnamefont {Avdeev}}, \bibinfo {author} {\bibfnamefont
  {E.}~\bibnamefont {Suard}}, \bibinfo {author} {\bibfnamefont {S.~A.~J.}\
  \bibnamefont {Kimber}}, \bibinfo {author} {\bibfnamefont {D.}~\bibnamefont
  {Lan\ifmmode~\mbox{\c{c}}\else \c{c}\fi{}on}}, \bibinfo {author}
  {\bibfnamefont {G.}~\bibnamefont {Pepe}}, \bibinfo {author} {\bibfnamefont
  {B.}~\bibnamefont {Moubaraki}}, \ and\ \bibinfo {author} {\bibfnamefont
  {T.~J.}\ \bibnamefont {Hicks}},\ }\href
  {https://link.aps.org/doi/10.1103/PhysRevB.92.224408} {\bibfield  {journal}
  {\bibinfo  {journal} {Phys. Rev. B}\ }\textbf {\bibinfo {volume} {92}},\
  \bibinfo {pages} {224408} (\bibinfo {year} {2015})}\BibitemShut {NoStop}%
\bibitem [{\citenamefont {Kim}\ \emph {et~al.}(2019{\natexlab{b}})\citenamefont
  {Kim}, \citenamefont {Lim}, \citenamefont {Lee}, \citenamefont {Lee},
  \citenamefont {Kim}, \citenamefont {Park}, \citenamefont {Jeon},
  \citenamefont {Park}, \citenamefont {Park},\ and\ \citenamefont
  {Cheong}}]{s41467-018-08284-6}%
  \BibitemOpen
  \bibfield  {author} {\bibinfo {author} {\bibfnamefont {K.}~\bibnamefont
  {Kim}}, \bibinfo {author} {\bibfnamefont {S.~Y.}\ \bibnamefont {Lim}},
  \bibinfo {author} {\bibfnamefont {J.-U.}\ \bibnamefont {Lee}}, \bibinfo
  {author} {\bibfnamefont {S.}~\bibnamefont {Lee}}, \bibinfo {author}
  {\bibfnamefont {T.~Y.}\ \bibnamefont {Kim}}, \bibinfo {author} {\bibfnamefont
  {K.}~\bibnamefont {Park}}, \bibinfo {author} {\bibfnamefont {G.~S.}\
  \bibnamefont {Jeon}}, \bibinfo {author} {\bibfnamefont {C.-H.}\ \bibnamefont
  {Park}}, \bibinfo {author} {\bibfnamefont {J.-G.}\ \bibnamefont {Park}}, \
  and\ \bibinfo {author} {\bibfnamefont {H.}~\bibnamefont {Cheong}},\ }\href
  {https://doi.org/10.1038/s41467-018-08284-6} {\bibfield  {journal} {\bibinfo
  {journal} {Nat. Commun.}\ }\textbf {\bibinfo {volume} {10}},\ \bibinfo
  {pages} {345} (\bibinfo {year} {2019}{\natexlab{b}})}\BibitemShut {NoStop}%
\bibitem [{\citenamefont {Johnson}\ \emph {et~al.}(2015)\citenamefont
  {Johnson}, \citenamefont {Williams}, \citenamefont {Haghighirad},
  \citenamefont {Singleton}, \citenamefont {Zapf}, \citenamefont {Manuel},
  \citenamefont {Mazin}, \citenamefont {Li}, \citenamefont {Jeschke},
  \citenamefont {Valent\'{\i}},\ and\ \citenamefont
  {Coldea}}]{PhysRevB.92.235119}%
  \BibitemOpen
  \bibfield  {author} {\bibinfo {author} {\bibfnamefont {R.~D.}\ \bibnamefont
  {Johnson}}, \bibinfo {author} {\bibfnamefont {S.~C.}\ \bibnamefont
  {Williams}}, \bibinfo {author} {\bibfnamefont {A.~A.}\ \bibnamefont
  {Haghighirad}}, \bibinfo {author} {\bibfnamefont {J.}~\bibnamefont
  {Singleton}}, \bibinfo {author} {\bibfnamefont {V.}~\bibnamefont {Zapf}},
  \bibinfo {author} {\bibfnamefont {P.}~\bibnamefont {Manuel}}, \bibinfo
  {author} {\bibfnamefont {I.~I.}\ \bibnamefont {Mazin}}, \bibinfo {author}
  {\bibfnamefont {Y.}~\bibnamefont {Li}}, \bibinfo {author} {\bibfnamefont
  {H.~O.}\ \bibnamefont {Jeschke}}, \bibinfo {author} {\bibfnamefont
  {R.}~\bibnamefont {Valent\'{\i}}}, \ and\ \bibinfo {author} {\bibfnamefont
  {R.}~\bibnamefont {Coldea}},\ }\href
  {https://link.aps.org/doi/10.1103/PhysRevB.92.235119} {\bibfield  {journal}
  {\bibinfo  {journal} {Phys. Rev. B}\ }\textbf {\bibinfo {volume} {92}},\
  \bibinfo {pages} {235119} (\bibinfo {year} {2015})}\BibitemShut {NoStop}%
\bibitem [{\citenamefont {Chaloupka}\ \emph {et~al.}(2013)\citenamefont
  {Chaloupka}, \citenamefont {Jackeli},\ and\ \citenamefont
  {Khaliullin}}]{PhysRevLett.110.097204}%
  \BibitemOpen
  \bibfield  {author} {\bibinfo {author} {\bibfnamefont {J.~c.~v.}\
  \bibnamefont {Chaloupka}}, \bibinfo {author} {\bibfnamefont {G.}~\bibnamefont
  {Jackeli}}, \ and\ \bibinfo {author} {\bibfnamefont {G.}~\bibnamefont
  {Khaliullin}},\ }\href
  {https://link.aps.org/doi/10.1103/PhysRevLett.110.097204} {\bibfield
  {journal} {\bibinfo  {journal} {Phys. Rev. Lett.}\ }\textbf {\bibinfo
  {volume} {110}},\ \bibinfo {pages} {097204} (\bibinfo {year}
  {2013})}\BibitemShut {NoStop}%
\bibitem [{\citenamefont {Cenker}\ \emph {et~al.}(2021)\citenamefont {Cenker},
  \citenamefont {Huang}, \citenamefont {Suri}, \citenamefont {Thijssen},
  \citenamefont {Miller}, \citenamefont {Song}, \citenamefont {Taniguchi},
  \citenamefont {Watanabe}, \citenamefont {McGuire}, \citenamefont {Xiao},\
  and\ \citenamefont {Xu}}]{s41567-020-0999-1}%
  \BibitemOpen
  \bibfield  {author} {\bibinfo {author} {\bibfnamefont {J.}~\bibnamefont
  {Cenker}}, \bibinfo {author} {\bibfnamefont {B.}~\bibnamefont {Huang}},
  \bibinfo {author} {\bibfnamefont {N.}~\bibnamefont {Suri}}, \bibinfo {author}
  {\bibfnamefont {P.}~\bibnamefont {Thijssen}}, \bibinfo {author}
  {\bibfnamefont {A.}~\bibnamefont {Miller}}, \bibinfo {author} {\bibfnamefont
  {T.}~\bibnamefont {Song}}, \bibinfo {author} {\bibfnamefont {T.}~\bibnamefont
  {Taniguchi}}, \bibinfo {author} {\bibfnamefont {K.}~\bibnamefont {Watanabe}},
  \bibinfo {author} {\bibfnamefont {M.~A.}\ \bibnamefont {McGuire}}, \bibinfo
  {author} {\bibfnamefont {D.}~\bibnamefont {Xiao}}, \ and\ \bibinfo {author}
  {\bibfnamefont {X.}~\bibnamefont {Xu}},\ }\href
  {https://doi.org/10.1038/s41567-020-0999-1} {\bibfield  {journal} {\bibinfo
  {journal} {Nat. Phys.}\ }\textbf {\bibinfo {volume} {17}},\ \bibinfo {pages}
  {20} (\bibinfo {year} {2021})}\BibitemShut {NoStop}%
\bibitem [{\citenamefont {Liu}\ and\ \citenamefont
  {Petrovic}(2018)}]{PhysRevB.97.014420}%
  \BibitemOpen
  \bibfield  {author} {\bibinfo {author} {\bibfnamefont {Y.}~\bibnamefont
  {Liu}}\ and\ \bibinfo {author} {\bibfnamefont {C.}~\bibnamefont {Petrovic}},\
  }\href {https://link.aps.org/doi/10.1103/PhysRevB.97.014420} {\bibfield
  {journal} {\bibinfo  {journal} {Phys. Rev. B}\ }\textbf {\bibinfo {volume}
  {97}},\ \bibinfo {pages} {014420} (\bibinfo {year} {2018})}\BibitemShut
  {NoStop}%
\bibitem [{\citenamefont {Holstein}\ and\ \citenamefont
  {Primakoff}(1940)}]{PhysRev.58.1098}%
  \BibitemOpen
  \bibfield  {author} {\bibinfo {author} {\bibfnamefont {T.}~\bibnamefont
  {Holstein}}\ and\ \bibinfo {author} {\bibfnamefont {H.}~\bibnamefont
  {Primakoff}},\ }\href {https://link.aps.org/doi/10.1103/PhysRev.58.1098}
  {\bibfield  {journal} {\bibinfo  {journal} {Phys. Rev.}\ }\textbf {\bibinfo
  {volume} {58}},\ \bibinfo {pages} {1098} (\bibinfo {year}
  {1940})}\BibitemShut {NoStop}%
\bibitem [{\citenamefont {Abramchuk}\ \emph {et~al.}(2018)\citenamefont
  {Abramchuk}, \citenamefont {Jaszewski}, \citenamefont {Metz}, \citenamefont
  {Osterhoudt}, \citenamefont {Wang}, \citenamefont {Burch},\ and\
  \citenamefont {Tafti}}]{adma.201801325}%
  \BibitemOpen
  \bibfield  {author} {\bibinfo {author} {\bibfnamefont {M.}~\bibnamefont
  {Abramchuk}}, \bibinfo {author} {\bibfnamefont {S.}~\bibnamefont
  {Jaszewski}}, \bibinfo {author} {\bibfnamefont {K.~R.}\ \bibnamefont {Metz}},
  \bibinfo {author} {\bibfnamefont {G.~B.}\ \bibnamefont {Osterhoudt}},
  \bibinfo {author} {\bibfnamefont {Y.}~\bibnamefont {Wang}}, \bibinfo {author}
  {\bibfnamefont {K.~S.}\ \bibnamefont {Burch}}, \ and\ \bibinfo {author}
  {\bibfnamefont {F.}~\bibnamefont {Tafti}},\ }\href
  {https://onlinelibrary.wiley.com/doi/abs/10.1002/adma.201801325} {\bibfield
  {journal} {\bibinfo  {journal} {Adv. Mater.}\ }\textbf {\bibinfo {volume}
  {30}},\ \bibinfo {pages} {1801325} (\bibinfo {year} {2018})}\BibitemShut
  {NoStop}%
\bibitem [{\citenamefont {Kim}\ \emph {et~al.}(2019{\natexlab{c}})\citenamefont
  {Kim}, \citenamefont {Kumaravadivel}, \citenamefont {Birkbeck}, \citenamefont
  {Kuang}, \citenamefont {Xu}, \citenamefont {Hopkinson}, \citenamefont
  {Knolle}, \citenamefont {McClarty}, \citenamefont {Berdyugin}, \citenamefont
  {Ben~Shalom}, \citenamefont {Gorbachev}, \citenamefont {Haigh}, \citenamefont
  {Liu}, \citenamefont {Edgar}, \citenamefont {Novoselov}, \citenamefont
  {Grigorieva},\ and\ \citenamefont {Geim}}]{s41928-019-0302-6}%
  \BibitemOpen
  \bibfield  {author} {\bibinfo {author} {\bibfnamefont {M.}~\bibnamefont
  {Kim}}, \bibinfo {author} {\bibfnamefont {P.}~\bibnamefont {Kumaravadivel}},
  \bibinfo {author} {\bibfnamefont {J.}~\bibnamefont {Birkbeck}}, \bibinfo
  {author} {\bibfnamefont {W.}~\bibnamefont {Kuang}}, \bibinfo {author}
  {\bibfnamefont {S.~G.}\ \bibnamefont {Xu}}, \bibinfo {author} {\bibfnamefont
  {D.~G.}\ \bibnamefont {Hopkinson}}, \bibinfo {author} {\bibfnamefont
  {J.}~\bibnamefont {Knolle}}, \bibinfo {author} {\bibfnamefont {P.~A.}\
  \bibnamefont {McClarty}}, \bibinfo {author} {\bibfnamefont {A.~I.}\
  \bibnamefont {Berdyugin}}, \bibinfo {author} {\bibfnamefont {M.}~\bibnamefont
  {Ben~Shalom}}, \bibinfo {author} {\bibfnamefont {R.~V.}\ \bibnamefont
  {Gorbachev}}, \bibinfo {author} {\bibfnamefont {S.~J.}\ \bibnamefont
  {Haigh}}, \bibinfo {author} {\bibfnamefont {S.}~\bibnamefont {Liu}}, \bibinfo
  {author} {\bibfnamefont {J.~H.}\ \bibnamefont {Edgar}}, \bibinfo {author}
  {\bibfnamefont {K.~S.}\ \bibnamefont {Novoselov}}, \bibinfo {author}
  {\bibfnamefont {I.~V.}\ \bibnamefont {Grigorieva}}, \ and\ \bibinfo {author}
  {\bibfnamefont {A.~K.}\ \bibnamefont {Geim}},\ }\href
  {https://doi.org/10.1038/s41928-019-0302-6} {\bibfield  {journal} {\bibinfo
  {journal} {Nat. Electron.}\ }\textbf {\bibinfo {volume} {2}},\ \bibinfo
  {pages} {457} (\bibinfo {year} {2019}{\natexlab{c}})}\BibitemShut {NoStop}%
\bibitem [{\citenamefont {Tsubokawa}(1960)}]{JPSJ.15.1664}%
  \BibitemOpen
  \bibfield  {author} {\bibinfo {author} {\bibfnamefont {I.}~\bibnamefont
  {Tsubokawa}},\ }\href {https://doi.org/10.1143/JPSJ.15.1664} {\bibfield
  {journal} {\bibinfo  {journal} {J. Phys. Soc. Jpn.}\ }\textbf {\bibinfo
  {volume} {15}},\ \bibinfo {pages} {1664} (\bibinfo {year}
  {1960})}\BibitemShut {NoStop}%
\bibitem [{\citenamefont {Xue}\ \emph {et~al.}(2019)\citenamefont {Xue},
  \citenamefont {Hou}, \citenamefont {Wang},\ and\ \citenamefont
  {Wu}}]{PhysRevB.100.224429}%
  \BibitemOpen
  \bibfield  {author} {\bibinfo {author} {\bibfnamefont {F.}~\bibnamefont
  {Xue}}, \bibinfo {author} {\bibfnamefont {Y.}~\bibnamefont {Hou}}, \bibinfo
  {author} {\bibfnamefont {Z.}~\bibnamefont {Wang}}, \ and\ \bibinfo {author}
  {\bibfnamefont {R.}~\bibnamefont {Wu}},\ }\href
  {https://link.aps.org/doi/10.1103/PhysRevB.100.224429} {\bibfield  {journal}
  {\bibinfo  {journal} {Phys. Rev. B}\ }\textbf {\bibinfo {volume} {100}},\
  \bibinfo {pages} {224429} (\bibinfo {year} {2019})}\BibitemShut {NoStop}%
\bibitem [{\citenamefont {Lin}\ \emph {et~al.}(2017)\citenamefont {Lin},
  \citenamefont {Zhuang}, \citenamefont {Luo}, \citenamefont {Liu},
  \citenamefont {Chen}, \citenamefont {Yan}, \citenamefont {Sun}, \citenamefont
  {Zhou}, \citenamefont {Lu}, \citenamefont {Tong}, \citenamefont {Sheng},
  \citenamefont {Qu}, \citenamefont {Song}, \citenamefont {Zhu},\ and\
  \citenamefont {Sun}}]{PhysRevB.95.245212}%
  \BibitemOpen
  \bibfield  {author} {\bibinfo {author} {\bibfnamefont {G.~T.}\ \bibnamefont
  {Lin}}, \bibinfo {author} {\bibfnamefont {H.~L.}\ \bibnamefont {Zhuang}},
  \bibinfo {author} {\bibfnamefont {X.}~\bibnamefont {Luo}}, \bibinfo {author}
  {\bibfnamefont {B.~J.}\ \bibnamefont {Liu}}, \bibinfo {author} {\bibfnamefont
  {F.~C.}\ \bibnamefont {Chen}}, \bibinfo {author} {\bibfnamefont
  {J.}~\bibnamefont {Yan}}, \bibinfo {author} {\bibfnamefont {Y.}~\bibnamefont
  {Sun}}, \bibinfo {author} {\bibfnamefont {J.}~\bibnamefont {Zhou}}, \bibinfo
  {author} {\bibfnamefont {W.~J.}\ \bibnamefont {Lu}}, \bibinfo {author}
  {\bibfnamefont {P.}~\bibnamefont {Tong}}, \bibinfo {author} {\bibfnamefont
  {Z.~G.}\ \bibnamefont {Sheng}}, \bibinfo {author} {\bibfnamefont
  {Z.}~\bibnamefont {Qu}}, \bibinfo {author} {\bibfnamefont {W.~H.}\
  \bibnamefont {Song}}, \bibinfo {author} {\bibfnamefont {X.~B.}\ \bibnamefont
  {Zhu}}, \ and\ \bibinfo {author} {\bibfnamefont {Y.~P.}\ \bibnamefont
  {Sun}},\ }\href {https://link.aps.org/doi/10.1103/PhysRevB.95.245212}
  {\bibfield  {journal} {\bibinfo  {journal} {Phys. Rev. B}\ }\textbf {\bibinfo
  {volume} {95}},\ \bibinfo {pages} {245212} (\bibinfo {year}
  {2017})}\BibitemShut {NoStop}%
\bibitem [{\citenamefont {Liu}\ \emph {et~al.}(2018)\citenamefont {Liu},
  \citenamefont {Dai}, \citenamefont {Yang}, \citenamefont {Fan}, \citenamefont
  {Pi}, \citenamefont {Zhang},\ and\ \citenamefont
  {Zhang}}]{PhysRevB.98.214420}%
  \BibitemOpen
  \bibfield  {author} {\bibinfo {author} {\bibfnamefont {W.}~\bibnamefont
  {Liu}}, \bibinfo {author} {\bibfnamefont {Y.}~\bibnamefont {Dai}}, \bibinfo
  {author} {\bibfnamefont {Y.-E.}\ \bibnamefont {Yang}}, \bibinfo {author}
  {\bibfnamefont {J.}~\bibnamefont {Fan}}, \bibinfo {author} {\bibfnamefont
  {L.}~\bibnamefont {Pi}}, \bibinfo {author} {\bibfnamefont {L.}~\bibnamefont
  {Zhang}}, \ and\ \bibinfo {author} {\bibfnamefont {Y.}~\bibnamefont
  {Zhang}},\ }\href {https://link.aps.org/doi/10.1103/PhysRevB.98.214420}
  {\bibfield  {journal} {\bibinfo  {journal} {Phys. Rev. B}\ }\textbf {\bibinfo
  {volume} {98}},\ \bibinfo {pages} {214420} (\bibinfo {year}
  {2018})}\BibitemShut {NoStop}%
\bibitem [{\citenamefont {Khan}\ \emph {et~al.}(2019)\citenamefont {Khan},
  \citenamefont {Zollitsch}, \citenamefont {Arroo}, \citenamefont {Cheng},
  \citenamefont {Verzhbitskiy}, \citenamefont {Sud}, \citenamefont {Feng},
  \citenamefont {Eda},\ and\ \citenamefont
  {Kurebayashi}}]{PhysRevB.100.134437}%
  \BibitemOpen
  \bibfield  {author} {\bibinfo {author} {\bibfnamefont {S.}~\bibnamefont
  {Khan}}, \bibinfo {author} {\bibfnamefont {C.~W.}\ \bibnamefont {Zollitsch}},
  \bibinfo {author} {\bibfnamefont {D.~M.}\ \bibnamefont {Arroo}}, \bibinfo
  {author} {\bibfnamefont {H.}~\bibnamefont {Cheng}}, \bibinfo {author}
  {\bibfnamefont {I.}~\bibnamefont {Verzhbitskiy}}, \bibinfo {author}
  {\bibfnamefont {A.}~\bibnamefont {Sud}}, \bibinfo {author} {\bibfnamefont
  {Y.~P.}\ \bibnamefont {Feng}}, \bibinfo {author} {\bibfnamefont
  {G.}~\bibnamefont {Eda}}, \ and\ \bibinfo {author} {\bibfnamefont
  {H.}~\bibnamefont {Kurebayashi}},\ }\href
  {https://link.aps.org/doi/10.1103/PhysRevB.100.134437} {\bibfield  {journal}
  {\bibinfo  {journal} {Phys. Rev. B}\ }\textbf {\bibinfo {volume} {100}},\
  \bibinfo {pages} {134437} (\bibinfo {year} {2019})}\BibitemShut {NoStop}%
\bibitem [{\citenamefont {Wang}\ \emph {et~al.}(2018)\citenamefont {Wang},
  \citenamefont {Zhang}, \citenamefont {Ding}, \citenamefont {Dong},
  \citenamefont {Li}, \citenamefont {Chen}, \citenamefont {Li}, \citenamefont
  {Huang}, \citenamefont {Wang}, \citenamefont {Zhao}, \citenamefont {Li},
  \citenamefont {Li}, \citenamefont {Jia}, \citenamefont {Sun}, \citenamefont
  {Guo}, \citenamefont {Ye}, \citenamefont {Sun}, \citenamefont {Chen},
  \citenamefont {Yang}, \citenamefont {Zhang}, \citenamefont {Ono},
  \citenamefont {Han},\ and\ \citenamefont {Zhang}}]{s41565-018-0186-z}%
  \BibitemOpen
  \bibfield  {author} {\bibinfo {author} {\bibfnamefont {Z.}~\bibnamefont
  {Wang}}, \bibinfo {author} {\bibfnamefont {T.}~\bibnamefont {Zhang}},
  \bibinfo {author} {\bibfnamefont {M.}~\bibnamefont {Ding}}, \bibinfo {author}
  {\bibfnamefont {B.}~\bibnamefont {Dong}}, \bibinfo {author} {\bibfnamefont
  {Y.}~\bibnamefont {Li}}, \bibinfo {author} {\bibfnamefont {M.}~\bibnamefont
  {Chen}}, \bibinfo {author} {\bibfnamefont {X.}~\bibnamefont {Li}}, \bibinfo
  {author} {\bibfnamefont {J.}~\bibnamefont {Huang}}, \bibinfo {author}
  {\bibfnamefont {H.}~\bibnamefont {Wang}}, \bibinfo {author} {\bibfnamefont
  {X.}~\bibnamefont {Zhao}}, \bibinfo {author} {\bibfnamefont {Y.}~\bibnamefont
  {Li}}, \bibinfo {author} {\bibfnamefont {D.}~\bibnamefont {Li}}, \bibinfo
  {author} {\bibfnamefont {C.}~\bibnamefont {Jia}}, \bibinfo {author}
  {\bibfnamefont {L.}~\bibnamefont {Sun}}, \bibinfo {author} {\bibfnamefont
  {H.}~\bibnamefont {Guo}}, \bibinfo {author} {\bibfnamefont {Y.}~\bibnamefont
  {Ye}}, \bibinfo {author} {\bibfnamefont {D.}~\bibnamefont {Sun}}, \bibinfo
  {author} {\bibfnamefont {Y.}~\bibnamefont {Chen}}, \bibinfo {author}
  {\bibfnamefont {T.}~\bibnamefont {Yang}}, \bibinfo {author} {\bibfnamefont
  {J.}~\bibnamefont {Zhang}}, \bibinfo {author} {\bibfnamefont
  {S.}~\bibnamefont {Ono}}, \bibinfo {author} {\bibfnamefont {Z.}~\bibnamefont
  {Han}}, \ and\ \bibinfo {author} {\bibfnamefont {Z.}~\bibnamefont {Zhang}},\
  }\href {https://doi.org/10.1038/s41565-018-0186-z} {\bibfield  {journal}
  {\bibinfo  {journal} {Nat. Nanotechnol.}\ }\textbf {\bibinfo {volume} {13}},\
  \bibinfo {pages} {554} (\bibinfo {year} {2018})}\BibitemShut {NoStop}%
\bibitem [{\citenamefont {Jiang}\ \emph {et~al.}(2018)\citenamefont {Jiang},
  \citenamefont {Li}, \citenamefont {Wang}, \citenamefont {Mak},\ and\
  \citenamefont {Shan}}]{s41565-018-0135-x}%
  \BibitemOpen
  \bibfield  {author} {\bibinfo {author} {\bibfnamefont {S.}~\bibnamefont
  {Jiang}}, \bibinfo {author} {\bibfnamefont {L.}~\bibnamefont {Li}}, \bibinfo
  {author} {\bibfnamefont {Z.}~\bibnamefont {Wang}}, \bibinfo {author}
  {\bibfnamefont {K.~F.}\ \bibnamefont {Mak}}, \ and\ \bibinfo {author}
  {\bibfnamefont {J.}~\bibnamefont {Shan}},\ }\href
  {https://doi.org/10.1038/s41565-018-0135-x} {\bibfield  {journal} {\bibinfo
  {journal} {Nat. Nanotechnol.}\ }\textbf {\bibinfo {volume} {13}},\ \bibinfo
  {pages} {549} (\bibinfo {year} {2018})}\BibitemShut {NoStop}%
\bibitem [{\citenamefont {Matsukura}\ \emph {et~al.}(2015)\citenamefont
  {Matsukura}, \citenamefont {Tokura},\ and\ \citenamefont
  {Ohno}}]{nnano.2015.22}%
  \BibitemOpen
  \bibfield  {author} {\bibinfo {author} {\bibfnamefont {F.}~\bibnamefont
  {Matsukura}}, \bibinfo {author} {\bibfnamefont {Y.}~\bibnamefont {Tokura}}, \
  and\ \bibinfo {author} {\bibfnamefont {H.}~\bibnamefont {Ohno}},\ }\href
  {https://doi.org/10.1038/nnano.2015.22} {\bibfield  {journal} {\bibinfo
  {journal} {Nat. Nanotechnol.}\ }\textbf {\bibinfo {volume} {10}},\ \bibinfo
  {pages} {209} (\bibinfo {year} {2015})}\BibitemShut {NoStop}%
\bibitem [{\citenamefont {Huang}\ \emph {et~al.}(2018)\citenamefont {Huang},
  \citenamefont {Clark}, \citenamefont {Klein}, \citenamefont {MacNeill},
  \citenamefont {Navarro-Moratalla}, \citenamefont {Seyler}, \citenamefont
  {Wilson}, \citenamefont {McGuire}, \citenamefont {Cobden}, \citenamefont
  {Xiao}, \citenamefont {Yao}, \citenamefont {Jarillo-Herrero},\ and\
  \citenamefont {Xu}}]{s41565-018-0121-3}%
  \BibitemOpen
  \bibfield  {author} {\bibinfo {author} {\bibfnamefont {B.}~\bibnamefont
  {Huang}}, \bibinfo {author} {\bibfnamefont {G.}~\bibnamefont {Clark}},
  \bibinfo {author} {\bibfnamefont {D.~R.}\ \bibnamefont {Klein}}, \bibinfo
  {author} {\bibfnamefont {D.}~\bibnamefont {MacNeill}}, \bibinfo {author}
  {\bibfnamefont {E.}~\bibnamefont {Navarro-Moratalla}}, \bibinfo {author}
  {\bibfnamefont {K.~L.}\ \bibnamefont {Seyler}}, \bibinfo {author}
  {\bibfnamefont {N.}~\bibnamefont {Wilson}}, \bibinfo {author} {\bibfnamefont
  {M.~A.}\ \bibnamefont {McGuire}}, \bibinfo {author} {\bibfnamefont {D.~H.}\
  \bibnamefont {Cobden}}, \bibinfo {author} {\bibfnamefont {D.}~\bibnamefont
  {Xiao}}, \bibinfo {author} {\bibfnamefont {W.}~\bibnamefont {Yao}}, \bibinfo
  {author} {\bibfnamefont {P.}~\bibnamefont {Jarillo-Herrero}}, \ and\ \bibinfo
  {author} {\bibfnamefont {X.}~\bibnamefont {Xu}},\ }\href
  {https://doi.org/10.1038/s41565-018-0121-3} {\bibfield  {journal} {\bibinfo
  {journal} {Nat. Nanotechnol.}\ }\textbf {\bibinfo {volume} {13}},\ \bibinfo
  {pages} {544} (\bibinfo {year} {2018})}\BibitemShut {NoStop}%
\bibitem [{\citenamefont {Biscaras}\ \emph {et~al.}(2015)\citenamefont
  {Biscaras}, \citenamefont {Chen}, \citenamefont {Paradisi},\ and\
  \citenamefont {Shukla}}]{ncomms9826}%
  \BibitemOpen
  \bibfield  {author} {\bibinfo {author} {\bibfnamefont {J.}~\bibnamefont
  {Biscaras}}, \bibinfo {author} {\bibfnamefont {Z.}~\bibnamefont {Chen}},
  \bibinfo {author} {\bibfnamefont {A.}~\bibnamefont {Paradisi}}, \ and\
  \bibinfo {author} {\bibfnamefont {A.}~\bibnamefont {Shukla}},\ }\href
  {https://doi.org/10.1038/ncomms9826} {\bibfield  {journal} {\bibinfo
  {journal} {Nat. Commun.}\ }\textbf {\bibinfo {volume} {6}},\ \bibinfo {pages}
  {8826} (\bibinfo {year} {2015})}\BibitemShut {NoStop}%
\bibitem [{\citenamefont {Costanzo}\ \emph {et~al.}(2016)\citenamefont
  {Costanzo}, \citenamefont {Jo}, \citenamefont {Berger},\ and\ \citenamefont
  {Morpurgo}}]{nnano.2015.314}%
  \BibitemOpen
  \bibfield  {author} {\bibinfo {author} {\bibfnamefont {D.}~\bibnamefont
  {Costanzo}}, \bibinfo {author} {\bibfnamefont {S.}~\bibnamefont {Jo}},
  \bibinfo {author} {\bibfnamefont {H.}~\bibnamefont {Berger}}, \ and\ \bibinfo
  {author} {\bibfnamefont {A.~F.}\ \bibnamefont {Morpurgo}},\ }\href
  {https://doi.org/10.1038/nnano.2015.314} {\bibfield  {journal} {\bibinfo
  {journal} {Nat. Nanotechnol.}\ }\textbf {\bibinfo {volume} {11}},\ \bibinfo
  {pages} {339} (\bibinfo {year} {2016})}\BibitemShut {NoStop}%
\bibitem [{\citenamefont {Liu}\ \emph {et~al.}(2020)\citenamefont {Liu},
  \citenamefont {Liu}, \citenamefont {Yang}, \citenamefont {Chen},
  \citenamefont {Zhang}, \citenamefont {Li}, \citenamefont {Wu}, \citenamefont
  {Ruan}, \citenamefont {Xiu}, \citenamefont {Liu}, \citenamefont {He},
  \citenamefont {Zhang},\ and\ \citenamefont {Xu}}]{PhysRevLett.125.267205}%
  \BibitemOpen
  \bibfield  {author} {\bibinfo {author} {\bibfnamefont {B.}~\bibnamefont
  {Liu}}, \bibinfo {author} {\bibfnamefont {S.}~\bibnamefont {Liu}}, \bibinfo
  {author} {\bibfnamefont {L.}~\bibnamefont {Yang}}, \bibinfo {author}
  {\bibfnamefont {Z.}~\bibnamefont {Chen}}, \bibinfo {author} {\bibfnamefont
  {E.}~\bibnamefont {Zhang}}, \bibinfo {author} {\bibfnamefont
  {Z.}~\bibnamefont {Li}}, \bibinfo {author} {\bibfnamefont {J.}~\bibnamefont
  {Wu}}, \bibinfo {author} {\bibfnamefont {X.}~\bibnamefont {Ruan}}, \bibinfo
  {author} {\bibfnamefont {F.}~\bibnamefont {Xiu}}, \bibinfo {author}
  {\bibfnamefont {W.}~\bibnamefont {Liu}}, \bibinfo {author} {\bibfnamefont
  {L.}~\bibnamefont {He}}, \bibinfo {author} {\bibfnamefont {R.}~\bibnamefont
  {Zhang}}, \ and\ \bibinfo {author} {\bibfnamefont {Y.}~\bibnamefont {Xu}},\
  }\href {https://link.aps.org/doi/10.1103/PhysRevLett.125.267205} {\bibfield
  {journal} {\bibinfo  {journal} {Phys. Rev. Lett.}\ }\textbf {\bibinfo
  {volume} {125}},\ \bibinfo {pages} {267205} (\bibinfo {year}
  {2020})}\BibitemShut {NoStop}%
\bibitem [{\citenamefont {Song}\ \emph {et~al.}(2019)\citenamefont {Song},
  \citenamefont {Fei}, \citenamefont {Yankowitz}, \citenamefont {Lin},
  \citenamefont {Jiang}, \citenamefont {Hwangbo}, \citenamefont {Zhang},
  \citenamefont {Sun}, \citenamefont {Taniguchi}, \citenamefont {Watanabe},
  \citenamefont {McGuire}, \citenamefont {Graf}, \citenamefont {Cao},
  \citenamefont {Chu}, \citenamefont {Cobden}, \citenamefont {Dean},
  \citenamefont {Xiao},\ and\ \citenamefont {Xu}}]{s41563-019-0505-2}%
  \BibitemOpen
  \bibfield  {author} {\bibinfo {author} {\bibfnamefont {T.}~\bibnamefont
  {Song}}, \bibinfo {author} {\bibfnamefont {Z.}~\bibnamefont {Fei}}, \bibinfo
  {author} {\bibfnamefont {M.}~\bibnamefont {Yankowitz}}, \bibinfo {author}
  {\bibfnamefont {Z.}~\bibnamefont {Lin}}, \bibinfo {author} {\bibfnamefont
  {Q.}~\bibnamefont {Jiang}}, \bibinfo {author} {\bibfnamefont
  {K.}~\bibnamefont {Hwangbo}}, \bibinfo {author} {\bibfnamefont
  {Q.}~\bibnamefont {Zhang}}, \bibinfo {author} {\bibfnamefont
  {B.}~\bibnamefont {Sun}}, \bibinfo {author} {\bibfnamefont {T.}~\bibnamefont
  {Taniguchi}}, \bibinfo {author} {\bibfnamefont {K.}~\bibnamefont {Watanabe}},
  \bibinfo {author} {\bibfnamefont {M.~A.}\ \bibnamefont {McGuire}}, \bibinfo
  {author} {\bibfnamefont {D.}~\bibnamefont {Graf}}, \bibinfo {author}
  {\bibfnamefont {T.}~\bibnamefont {Cao}}, \bibinfo {author} {\bibfnamefont
  {J.-H.}\ \bibnamefont {Chu}}, \bibinfo {author} {\bibfnamefont {D.~H.}\
  \bibnamefont {Cobden}}, \bibinfo {author} {\bibfnamefont {C.~R.}\
  \bibnamefont {Dean}}, \bibinfo {author} {\bibfnamefont {D.}~\bibnamefont
  {Xiao}}, \ and\ \bibinfo {author} {\bibfnamefont {X.}~\bibnamefont {Xu}},\
  }\href {https://doi.org/10.1038/s41563-019-0505-2} {\bibfield  {journal}
  {\bibinfo  {journal} {Nat. Mater.}\ }\textbf {\bibinfo {volume} {18}},\
  \bibinfo {pages} {1298} (\bibinfo {year} {2019})}\BibitemShut {NoStop}%
\bibitem [{\citenamefont {Li}\ \emph {et~al.}(2019)\citenamefont {Li},
  \citenamefont {Jiang}, \citenamefont {Sivadas}, \citenamefont {Wang},
  \citenamefont {Xu}, \citenamefont {Weber}, \citenamefont {Goldberger},
  \citenamefont {Watanabe}, \citenamefont {Taniguchi}, \citenamefont {Fennie},
  \citenamefont {Fai~Mak},\ and\ \citenamefont {Shan}}]{s41563-019-0506-1}%
  \BibitemOpen
  \bibfield  {author} {\bibinfo {author} {\bibfnamefont {T.}~\bibnamefont
  {Li}}, \bibinfo {author} {\bibfnamefont {S.}~\bibnamefont {Jiang}}, \bibinfo
  {author} {\bibfnamefont {N.}~\bibnamefont {Sivadas}}, \bibinfo {author}
  {\bibfnamefont {Z.}~\bibnamefont {Wang}}, \bibinfo {author} {\bibfnamefont
  {Y.}~\bibnamefont {Xu}}, \bibinfo {author} {\bibfnamefont {D.}~\bibnamefont
  {Weber}}, \bibinfo {author} {\bibfnamefont {J.~E.}\ \bibnamefont
  {Goldberger}}, \bibinfo {author} {\bibfnamefont {K.}~\bibnamefont
  {Watanabe}}, \bibinfo {author} {\bibfnamefont {T.}~\bibnamefont {Taniguchi}},
  \bibinfo {author} {\bibfnamefont {C.~J.}\ \bibnamefont {Fennie}}, \bibinfo
  {author} {\bibfnamefont {K.}~\bibnamefont {Fai~Mak}}, \ and\ \bibinfo
  {author} {\bibfnamefont {J.}~\bibnamefont {Shan}},\ }\href
  {https://doi.org/10.1038/s41563-019-0506-1} {\bibfield  {journal} {\bibinfo
  {journal} {Nat. Mater.}\ }\textbf {\bibinfo {volume} {18}},\ \bibinfo {pages}
  {1303} (\bibinfo {year} {2019})}\BibitemShut {NoStop}%
\bibitem [{\citenamefont {Mondal}\ \emph {et~al.}(2019)\citenamefont {Mondal},
  \citenamefont {Kannan}, \citenamefont {Das}, \citenamefont {Govindaraj},
  \citenamefont {Singha}, \citenamefont {Satpati}, \citenamefont {Arumugam},\
  and\ \citenamefont {Mandal}}]{PhysRevB.99.180407}%
  \BibitemOpen
  \bibfield  {author} {\bibinfo {author} {\bibfnamefont {S.}~\bibnamefont
  {Mondal}}, \bibinfo {author} {\bibfnamefont {M.}~\bibnamefont {Kannan}},
  \bibinfo {author} {\bibfnamefont {M.}~\bibnamefont {Das}}, \bibinfo {author}
  {\bibfnamefont {L.}~\bibnamefont {Govindaraj}}, \bibinfo {author}
  {\bibfnamefont {R.}~\bibnamefont {Singha}}, \bibinfo {author} {\bibfnamefont
  {B.}~\bibnamefont {Satpati}}, \bibinfo {author} {\bibfnamefont
  {S.}~\bibnamefont {Arumugam}}, \ and\ \bibinfo {author} {\bibfnamefont
  {P.}~\bibnamefont {Mandal}},\ }\href
  {https://link.aps.org/doi/10.1103/PhysRevB.99.180407} {\bibfield  {journal}
  {\bibinfo  {journal} {Phys. Rev. B}\ }\textbf {\bibinfo {volume} {99}},\
  \bibinfo {pages} {180407} (\bibinfo {year} {2019})}\BibitemShut {NoStop}%
\bibitem [{\citenamefont {Liu}\ and\ \citenamefont
  {Petrovic}(2020)}]{PhysRevB.102.014424}%
  \BibitemOpen
  \bibfield  {author} {\bibinfo {author} {\bibfnamefont {Y.}~\bibnamefont
  {Liu}}\ and\ \bibinfo {author} {\bibfnamefont {C.}~\bibnamefont {Petrovic}},\
  }\href {https://link.aps.org/doi/10.1103/PhysRevB.102.014424} {\bibfield
  {journal} {\bibinfo  {journal} {Phys. Rev. B}\ }\textbf {\bibinfo {volume}
  {102}},\ \bibinfo {pages} {014424} (\bibinfo {year} {2020})}\BibitemShut
  {NoStop}%
\end{thebibliography}

%

\clearpage
\foreach \x in {1}
{%
\clearpage
\includepdf[pages={\x}]{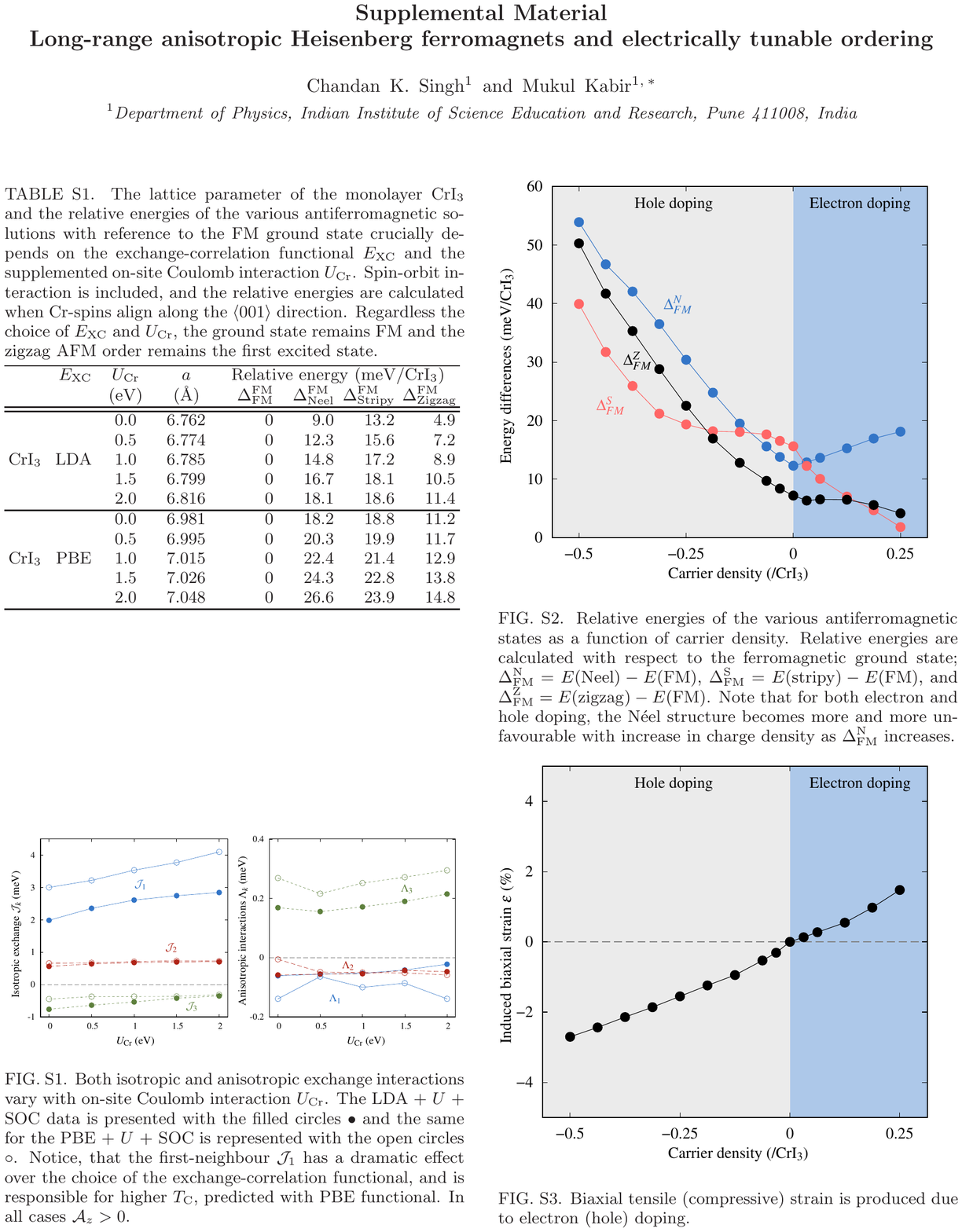}
}


\end{document}